\documentclass[11pt]{article} 

\usepackage[utf8]{inputenc} 

\usepackage{geometry} 
\usepackage{graphicx} 
\usepackage{xcolor}
\usepackage{parskip}
\usepackage{placeins}
\usepackage{hyperref}
\hypersetup{
    colorlinks=true,
    citecolor=cyan,
    linkcolor=.
}
\usepackage{bm}
\usepackage{amsmath}
\usepackage{amssymb}
\usepackage{subfig}
\usepackage{footnpag}
\usepackage{natbib}
\setcitestyle{authoryear,citesep={; },notesep={; },round,aysep={,},yysep={;}}
\usepackage{indentfirst}

\usepackage{lineno}
\definecolor{niceBlue}{rgb}{0.067, 0.404, 0.694}
\definecolor{niceRed}{rgb}{0.768, 0.118, 0.227}
\definecolor{nicePurple}{rgb}{0.28, 0.15, 0.77}

\title{Spin-Orbit Coupling of Europa's Ice Shell and Interior}
\author{%
Ethan R. Burnett\thanks{Laboratory for Atmospheric and Space Physics, University of Colorado Boulder, \texttt{ethan.burnett@colorado.edu}}, \ Paul O. Hayne\thanks{Astrophysical and Planetary Sciences; Laboratory for Atmospheric and Space Physics, University of Colorado Boulder, \texttt{paul.hayne@colorado.edu}}%
}

\date{} 
        
\begin{document}
\maketitle
\begin{abstract}
Europa is an icy ocean world, differentiated into a floating ice shell and solid interior, separated by a global ocean. The classical spin-orbit coupling problem considers a satellite as a single rigid body, but in the case of Europa, the existence of the subsurface ocean enables independent motion of the ice shell and solid interior. This paper explores the spin-orbit coupling problem for Europa from a dynamical perspective, yielding illuminating analytical and numerical results. We determine that the spin behavior of Europa is influenced by processes not captured by the classical single rigid body spin-orbit coupling analysis. The tidal locking process for Europa is governed by the strength of gravity-gradient coupling between the ice shell and solid interior, with qualitatively different behavior depending on the scale of this effect. In this coupled rigid model, the shell can potentially undergo large angular displacements from the solid interior, and the coupling plays an outsize role in the dynamical evolution of the moon, even without incorporating the dissipative effects of shell non-rigidity. We additionally discuss the effects of  a realistic viscoelastic shell, and catalogue other torques that we expect to be sub-dominant in Europa's spin dynamics, or whose importance is unknown. Finally, we explore how the choice of tidal model affects the resulting equilibrium spin state.
\end{abstract}

\section{Introduction}

The classical works of Goldreich and Peale in the 1960s showed that a planet in an eccentric orbit can settle into a spin rate $\omega_{p}$ that is slightly faster than the synchronous angular velocity $\omega_{p} = n$, where $n$ is the orbital mean motion \citep{Goldreich_TAJ1966,Goldreich_Peale_SpinOrb}. In the final spin state, the orbit-averaged tidal torque should equal zero, and this happens for a planetary spin rate greater the mean motion angular velocity and less the periapsis angular velocity. However, if the planet has sufficient permanent mass asymmetry (in addition to the dynamic tidal distortion), then gravity-gradient torques on this asymmetry can dominate the tidal torque and force synchronous rotation. Other resonant spin states e.g. $\omega_{p}/n = 1/2, 1, 3/2, 2, \ldots$ are also possible, but this paper is entirely focused on the synchronous case.

The analysis of \cite{Goldreich_TAJ1966} uses a 1D single-axis model of a single rigid body, and does not examine the case that the body is differentiated. The icy ocean world Europa is well-known to be differentiated into an ice shell and solid interior, dynamically decoupled by a subsurface ocean, and as a result, the simple conclusions of Goldreich and Peale will not apply. This is important because despite a wealth of relevant studies, it is still not known for certain whether or not Europa is in a state of super-synchronous rotation, nor is it known the degree to which its ice shell and core can move independently \citep{Ojakangas1989EuropaPW,Sarid2002PolarWander,GreenbergEuropaRotation2002,Schenk_TPW_circ,Burnett_Icarus}. Instead of using Goldreich and Peale's simple 1D model to approach the dynamics of this problem, a slightly more complex 2D model accounting for the independent single-axis rotational and librational movements of the ice shell and core is desirable. Such a model was previously developed by \cite{VanHoolst_Europa} to study librations of Europa's ice shell, but the model developed in that paper can also be applied to the larger problem of full rotations and tidal settling into synchronous or super-synchronous spin states. In this paper we explore the tidal locking process with this more appropriate model. We also discuss the unknown dynamical influences of the potentially highly dissipative ice shell and the global subsurface ocean. Additionally, we use previously derived tidal models whose accuracy potentially exceeds the simple but popular MacDonald model \citep{MacDonaldTidal,SSD_1999} used by Goldreich and Peale. Their use greatly reduces the appearance of the classical non-synchronous rotation solution in our numerical results. Altogether, this work calls into question the use of the classical spin-orbit coupling treatment for understanding Europa's present-day and historical spin states, as well as for the other icy ocean worlds. 

Europa has been a subject of intense interest since the Voyager missions, and this paper follows a long history of research into the orientation and spin history of Europa's icy surface. The Voyager mission revealed a smooth surface nearly devoid of impact craters, with a subsurface ocean suggested as the likely culprit for the rapid resurfacing \citep{EuropaVoyagerWater1983}. Due to the pervasive tectonic features observed on its surface, whose implied background stress field cannot be generated by diurnal tidal-driven stresses alone, it was suggested that Europa could rotate faster than the synchronous angular velocity, producing the necessary large stress fields in the ice shell \citep{Greenberg1984}.
Additional works have subsequently argued for non-synchronous rotation (NSR) by interpretation of tectonic features from both Voyager and Galileo data \citep{HoppaScience1999, Greenberg2003TidalStress,GreenbergEuropaRotation2002,Kattenhorn2002,Geissler1998}, suggesting very slow NSR periods on the order of $10^{4}$ to $10^{5}$ yrs. This corresponds to an unintuitive equilibrium state which appears synchronous on human timescales, but leaves its mark on the surface via a powerful stress field due to the strain from slow rotation over the unobservable millennia. It is worth noting that there are competing explanations for the source of the background stress field, including ice shell freezing \citep{Nimmo2004} and polar wander \citep{Greenberg2003TidalStress,Ojakangas1989EuropaPW,Sarid2002PolarWander}. Direct observation of NSR has proven quite difficult due to the extremely long expected timescale. \cite{Hoppa1999a} used pairs of images from Voyager 2 and by Galileo 17 years later to attempt a direct observation of NSR. They found no signature at all, and from the precision of their measurements, they determined that the period of NSR must be greater than 12000 years. Recently, \cite{Burnett_Icarus}  used Europa's hemispheric color dichotomy, which is produced by anisotropic exogenic processes, to search for a signature of NSR. We found no signature at all, and our analysis furthermore found that the synodic period of NSR should be at least $10^{6}$ yrs.

While many researchers have used observations from geology to argue for or against particular types of motion of Europa's ice shell, there is comparatively less literature focused on the study of Europa's overall spin state from a dynamical perspective. However, there are still some very important studies which were influential for this work. As previously mentioned, the model by \cite{VanHoolst_Europa} studied the gravitational coupling between Europa's ice shell and core, providing much of the theoretical groundwork for this study. The work of \cite{Ojakangas1989EuropaPW} is also quite relevant. They show that a decoupled ice shell on Europa can become dynamically unstable as it approaches thermal equilibrium, reorienting the shell by 90 degrees about the axis connecting the Jovian and sub-Jovian points. They also present a wealth of insights about the rotational dynamics of Europa's ice shell. Lastly, \cite{Bills_EuropaRotational} provides an authoritative study on the dynamics of Europa's ice shell, with a review of past work on the precession, libration, and potential non-synchronous rotation of Europa. They note that many of the relevant mechanisms and parameters are neither sufficiently well-measured nor well-understood for us to make convincing conclusions about Europa's spin state with dynamical arguments alone. For example, they derive an equilibrium NSR rate for Europa for which the (poorly constrained) tidal torque would vanish, and obtain a predicted synodic period of 15 years, noting that this result is far too short to be correct. Clearly the primary role of dynamical models for studying Europa 
is to aid understanding of the kinds of processes that could be at work there, and how they produce certain outcomes in Europa's dynamical evolution. It is to this end that this paper extends the classical spin-orbit coupling analysis of Goldreich and Peale to the differentiated icy ocean world Europa.

\section{Methods and Results}
The approach in this work is to examine how the analyses in \cite{Goldreich_TAJ1966} and \cite{Goldreich_Peale_SpinOrb} are extended to the case that a planet or satellite is differentiated into dynamically decoupled ice shell and core. In particular, it is of interest to see when full rotations of the ice shell, core, or both are dynamically permitted, and how the moments of inertia of the ice shell and core influence final body spin state behavior with tidal energy dissipation. This work begins with a brief review of the classical analysis.
\subsection{Dynamics of a Spinning Planet}
\begin{figure}[h!]
\centering
\includegraphics[width=2.8in]{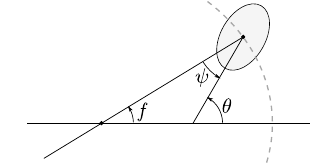}
\caption{Libration and spin angles for a rigid asymmetric satellite, where the asymmetry is depicted as an elongation of the planet. The satellite orbits about its focus with true anomaly $f$, rotation angle $\theta$, and libration angle $\psi$.}
\label{fig:goldreich1}
\end{figure}

Consider the situation depicted by Figure~\ref{fig:goldreich1}, discussed in \cite{Goldreich_TAJ1966}. A satellite with a persistent moss/moment of inertia asymmetry, depicted as an elongated axis, orbits with true anomaly $f$ and rotates with rotation angle $\theta$ measured from periapsis to the long axis of the planet. The angle $\psi$ is also defined, where $\psi = \theta - f$, and bounded libration occurs if $|\psi| < \pi/2$. 

From \cite{Goldreich_TAJ1966}, the dynamics of this problem are given below for the case of zero dynamic tidal torque:
\begin{equation}
\label{GoldEq1}
C\ddot{\psi} + \frac{3}{2}\frac{(B-A)GM}{r^{3}}\sin{2\psi} = -C\ddot{f}
\end{equation}
where $r$ is the orbit radius, $G$ is the gravitational constant, and $M$ is the mass of the attracting body. Neglecting the effects of eccentricity, the behavior of the averaged angle $\eta = \theta - nt$ is instructive. Its dynamics are given below:
\begin{equation}
\label{GoldEq2}
C\ddot{\eta} + \frac{3}{2}(B-A)n^{2}\sin{2\eta} = 0
\end{equation}
The angle $\psi$ behaves the same as $\eta$ but with additional small oscillations superimposed as a result of the orbital eccentricity. Eq.~\eqref{GoldEq2} is that of a nonlinear pendulum, and admits an energy integral:
\begin{equation}
\label{GoldEq3}
E = \frac{1}{2}C\dot{\eta}^{2} - \frac{3}{4}(B-A)n^{2}\cos{2\eta}
\end{equation}
While $E$ is only conserved for Eq.~\eqref{GoldEq2}, in the absence of the dynamic tidal torque (i.e. if the planet is perfectly rigid), the energy will oscillate only slightly about some conserved mean value for the system given by Eq.~\eqref{GoldEq1}. It turns out that the energy quantity is quite useful for many reasons, and this will be discussed later in the paper.

In contrast to the gravitational torque of the primary body on the permanent mass asymmetry of its satellite, the ``dynamic" tidal torque denotes the torque on the lagging asymmetric bulge raised by the tides. The dynamic tidal torque is written below using the MacDonald tidal assumption, which was originally made in Goldreich and Peale \citep{MacDonaldTidal,SSD_1999,Goldreich_TAJ1966}:
\begin{equation}
\label{TidesEq1} 
L_{\text{tidal}} = - \tilde{D}\left(\frac{a}{r}\right)^{6}\text{sign}(\dot{\psi})
\end{equation}
\begin{equation}
\label{TidesEq2}
\tilde{D} = \frac{3}{2}\frac{k_{2}}{Q_{s}}\frac{n^{4}}{G}R_{s}^{5}
\end{equation}
where $k_{2}$ is the second degree tidal Love number, $Q_{s}$ is the satellite's specific tidal dissipation factor, $G$ is the gravitational constant, and $R_{s}$ is the satellite radius. We defer to a discussion in Section 2.4.3 that the MacDonald tidal torque is fairly simplistic, and a use of the more complex Darwin-Kaula tidal torque model can be more reflective of real rheologies \citep{EfroimskyTidal}.

It can be shown that this tidal torque influences the total energy as below:
\begin{equation}
\label{GoldEq4}
\dot{E} = L_{\text{tidal}}(t)\dot{\eta} \ \approx \ \overline{L}_{\text{tidal}}\dot{\eta}
\end{equation}
where $\overline{L}_{\text{tidal}}$ is the orbit-averaged tidal torque, and changes to the energy are quite slow, so orbit averaging is appropriate. When the orbit-averaged tidal torque equals zero, there is no further change to the energy on average, and a final spin state is achieved. 
The investigation in \cite{Goldreich_TAJ1966} examines the long-term behavior by applying averaging to Eq.~\eqref{GoldEq4} obtaining the following well-known simple condition for a planet to stop rotating super-synchronously and settle into a state of libration:
\begin{equation}
\label{Ean2}
\left(\frac{3}{2}\left(\frac{B-A}{C}\right)\right)^{\frac{1}{2}} > \frac{9.5\pi e^{2}}{2\sqrt{2}}
\end{equation}
If this inequality is not satisfied, the planet will continue to rotate super-synchronously. Again note that this result rests on the use of the MacDonald tidal model. While quite popular, it is based on an analysis that is rheologically overly simplistic and mathematically incorrect \citep{EfroimskyMakarov2013} -- we discuss this in Section 2.4.3.

There are some challenges preventing an exact paralleling of Goldreich and Peale's energy-based analysis for the differentiated case. First, assuming significant independent motion is geometrically permitted, there are potentially two timescales at work - that of the libration/rotation of the ice shell, and the libration/rotation period of the core. Additionally, this problem presents two degrees of freedom, as opposed to the single degree-of-freedom analysis by Goldreich and Peale. We will show that even these simplified 2D dynamics can be chaotic, effectively precluding comprehensive analytic conclusions. However, there is still a lot of insight that can be obtained by deriving a new energy quantity and numerically exploring the effects of nonconservative torques.

\subsection{The Case of a Differentiated Body}
\begin{figure}[h!]
\centering
\includegraphics[]{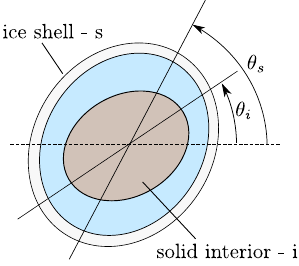}
\caption{An icy ocean world differentiated into an ice shell and solid interior, separated by a global ocean (not to scale). The angles $\theta_{i}$ and $\theta_{s}$ are inertial angles which measure the orientation of the permanent asymmetries of the solid interior and ice shell, respectively.}
\label{fig:EuropaInternal1}
\end{figure}

While the single rigid body assumption is suitable for spin-orbit coupling analysis of some worlds such as Mercury and the Moon, it is not appropriate for the icy ocean worlds, for which the complex interaction between ice shell and interior should not be discounted, and the shell itself can potentially become significantly displaced from the planetary interior on very short geologic timescales \citep{Ojakangas1989EuropaPW}. The insufficiency of the Goldreich \& Peale model is especially apparent for cases where the motion of the ice shell itself is of interest. In this work we in general consider a two-dimensional rotational model of Europa, depicted in Figure~\ref{fig:EuropaInternal1}. 

For a 2D spin-orbit coupling model of Europa, the generalized rotational dynamics can be expressed as:
\begin{subequations}
\label{Europa2Deq1}
\begin{align}
& C_{i}\ddot{\theta}_{i} = \Big(L_{\text{G,i}}(\psi_{i}) - L_{\text{C}}(\theta_{s}-\theta_{i}) \Big) + \delta L_{\text{i}}  \\
& C_{s}\ddot{\theta}_{s} = \Big(L_{\text{G,s}}(\psi_{s}) + L_{\text{C}}(\theta_{s}-\theta_{i}) \Big) + \delta L_{\text{s}} 
\end{align}
\end{subequations}
The inertially referenced angles $\theta_{i}$ and $\theta_{s}$ measure the orientation of the solid interior and ice shell, respectively. The Jovian-referenced libration angles are $\psi_{s} = \theta_{s} - f$ and $\psi_{i} = \theta_{i} - f$. Full revolutions of $\psi_{s}$ and $\psi_{i}$ only occur when the shell and interior respectively are not tidally locked. 
The terms $L_{G,s}$ and $L_{G,i}$ are the conservative Jovian gravitational torques on permanent asymmetries of the ice shell and solid interior, and $L_{C}$ denotes the sum of all conservative torques coupling the shell and interior. The terms $\delta L_{\text{s}}$ and $\delta L_{\text{i}}$ denote nonconservative torques. These include, among other things, both the dynamic tidal torque acting on the tidal bulge of the moon, and other dissipative torques, which are summarized later in this paper.

Modest orbit eccentricity induces small oscillations in $\psi_{s}$ and $\psi_{i}$, and the ``averaged" angles $\eta_{s} = \theta_{s} - nt$ and $\eta_{i} = \theta_{i} - nt$ are more convenient for analysis of non-synchronous rotation and tidal locking. To show this by transformation of Eq.~\eqref{Europa2Deq1}, note that $\ddot{\theta}_{j} = \ddot{\eta}_{j}$, and $\theta_{s} - \theta_{i} = \eta_{s} - \eta_{i}$. Furthermore $\psi_{j} = \eta_{j} - \varphi$ for differential orbit longitude $\varphi = f - nt = \mathcal{O}(e)$. For our purposes we are restricted to gravitational torques $L_{G,j}$ which can be expressed as $L_{G,j}(\psi_{j}) = L_{G,j}(\eta_{j}) + \delta L_{G,j}$ where $\delta L_{G,j}/L_{G,j}$ is of $\mathcal{O}(e)$. For example a sinusoidal term $\sin{(2\psi_{j})}$, akin to the gravitational torque term in Goldreich and Peale, reduces as below:
\begin{equation}
\label{sinReduce}
\begin{split}
\sin{(2\psi_{j})} & = \sin{(2\eta_{j})} - \sin{(2\varphi)}\cos{(2\eta_{j})} - \left(1 - \cos{(2\varphi)}\right)\sin{(2\eta_{j})} \\ & \approx \sin{(2\eta_{j})} - 2\varphi\cos{(2\eta_{j})} + \mathcal{O}(\delta\psi^{2}) \\
& = \sin{(2\eta_{j})} + \mathcal{O}(e)
\end{split}
\end{equation}
Thus Eq.~\eqref{Europa2Deq1} is transformed:
\begin{subequations}
\label{Europa2DavgTrans}
\begin{align}
& C_{i}\ddot{\eta}_{i} = \Big(L_{\text{G,i}}(\eta_{i}) - L_{\text{C}}(\eta_{s}-\eta_{i}) \Big) + \delta L_{\text{i}} + \delta L_{G,i} \\
& C_{s}\ddot{\eta}_{s} = \Big(L_{\text{G,s}}(\eta_{s}) + L_{\text{C}}(\eta_{s}-\eta_{i}) \Big) + \delta L_{\text{s}} + \delta L_{G,s} 
\end{align}
\end{subequations}
Neglecting the nonconservative disturbance torques, we recognize the remainder of Eq.~\eqref{Europa2DavgTrans} as a Hamiltonian system with conjugate angular momenta $\bm{p} = (C_{i}\dot{\eta}_{i}, C_{s}\dot{\eta}_{s})^{\top}$ and configuration coordinates $\bm{q} = \bm{\eta} = (\eta_{i}, \eta_{s})^{\top}$. Then the Hamiltonian $\mathcal{H} = \bm{p}^{\top}\dot{\bm{q}} - \mathcal{L}$ is given below, and applying Hamilton's equations we get the form that follows:
\begin{equation}
\label{GeneralHam}
\mathcal{H} = \frac{1}{2}C_{i}\dot{\eta}_{i}^{2} + \frac{1}{2}C_{s}\dot{\eta}_{s}^{2} + V(\bm{\eta})
\end{equation}
\begin{equation}
\label{GeneralHamEOM}
\frac{\text{d}}{\text{d}t}\begin{pmatrix} C_{i}\dot{\eta}_{i} \\ C_{s}\dot{\eta}_{s} \end{pmatrix} = -\begin{pmatrix} \frac{\partial}{\partial\eta_{i}}(V(\bm{\eta})) \\ \frac{\partial}{\partial\eta_{s}}(V(\bm{\eta})) \end{pmatrix}
\end{equation}
where the conservative torques are directly derived from the common scalar potential. Also, the scalar potential can be derived from any specified conservative torques using the relationships below: 
\begin{equation}
\label{VfromL1}
\begin{split}
V(\bm{\eta}) & = \int \Big( L_{\text{G,s}}(\eta_{s}) + L_{\text{C}}(\eta_{s}-\eta_{i}) \Big)\text{d}\eta_{s} + F(\eta_{i}) \\
& = \int \Big( L_{\text{G,i}}(\eta_{i}) - L_{\text{C}}(\eta_{s}-\eta_{i}) \Big)\text{d}\eta_{i} + F(\eta_{s})
\end{split}
\end{equation}

Note that for simplicity and convenience we will generally refer to the Hamiltonian integral $\mathcal{H}$ as the system energy or rotational energy $E$. We can show using the above formulation that the system energy is influenced by perturbative torques in the following way:
\begin{equation}
\label{EnergyDot}
\dot{E} = \dot{\mathcal{H}} = (\delta L_{i} + \delta L_{G,i})\dot{\eta}_{i} + (\delta L_{s} + \delta L_{G,s})\dot{\eta}_{s} 
\end{equation}
Orbit-averaging this result, we note the disappearance of the perturbations $\delta L_{G,i}$ due to eccentricity, which we assume orbit-average to zero. Eq.~\eqref{sinReduce} can be used to easily show that this is true to $\mathcal{O}(e)$ for the typical sinusoidal planetary gravitational torque term. What remains are the orbit-averaged nonconservative torques, including the dynamic tidal torques, and torques due to dissipation within the ice shell:
\begin{equation}
\label{EnergyDotAvg}
\dot{E} \approx \delta\overline{L}_{i}\dot{\eta}_{i} + \delta\overline{L}_{s}\dot{\eta}_{s} 
\end{equation}
An outline of major torques potentially at work in the spin-orbit coupling of icy moons is given later in the paper. 

To study how the total system energy relates to the permissible spin states, the concept of the \textit{zero-velocity curve} is borrowed from celestial mechanics -- see e.g. \cite{Koon:2006rf}. Namely, setting $\dot{\eta}_{s} = \dot{\eta}_{i} = 0$ and examining curves of constant values of $E$, for a given energy level, the range of reachable angles $\{\eta_{i}, \ \eta_{s} \}$ is externally bounded by the curve at that fixed energy value. We can explore this mathematically by noting for a specified energy level $E = E_{0}$, the kinetic energy is as below:
\begin{subequations}
\label{EtoZVC1}
\begin{align}
& T = E_{0} - V(\eta_{i},\eta_{s}) \\ 
& T = \frac{1}{2}C_{i}\dot{\eta}_{i}^{2} + \frac{1}{2}C_{s}\dot{\eta}_{s}^{2} \geq 0
\end{align}
\end{subequations}
The zero velocity curve is defined as the continuous equipotential curve of angles $\bm{\eta}~=~(\eta_{i},\eta_{s})^{\top}$ satisfying the relationship below:
\begin{equation}
\label{EtoZVC2}
\bm{\eta}^{*}(E_{0}) = \left\{\bm{\eta} : \left. E(\bm{\eta},\dot{\bm{\eta}})\right\vert_{\dot{\bm{\eta}}=0} = V(\bm{\eta}) = E_{0} \right\}
\end{equation}
Let states \textit{inside} the above zero-velocity curve be defined as $\bm{x}^{-}(E_{0})$:
\begin{equation}
\label{EtoZVC3}
\bm{x}^{-}(E_{0}) = \left\{\bm{\eta},\dot{\bm{\eta}} : V(\bm{\eta}) < E_{0} \ \text{and} \ E = T(\dot{\bm{\eta}}) + V(\bm{\eta}) = E_{0} \right\}
\end{equation}
For these states $T > 0$ and the angles will be in motion. By contrast, states \textit{outside} the zero-velocity curve $\bm{\eta}^{+}(E_{0})$ are defined below:
\begin{equation}
\label{EtoZVC4}
\bm{x}^{+}(E_{0}) = \left\{\bm{\eta},\dot{\bm{\eta}} : V(\bm{\eta}) > E_{0} \ \text{and} \ E = T(\dot{\bm{\eta}}) + V(\bm{\eta}) = E_{0} \right\}
\end{equation}
For these states, $T < 0$, which is unphysical, thus these outside states are dynamically prohibited. From this we can conclude that for a rotational state of energy level $E_{0}$, the configuration will be externally bounded for all time by the zero-velocity curve $\bm{\eta}^{*}(E_{0})$. With energy loss due to dissipation, this curve will contract and the accessible region of states will become increasingly limited. 

\begin{figure}[h!]
\centering
\includegraphics[]{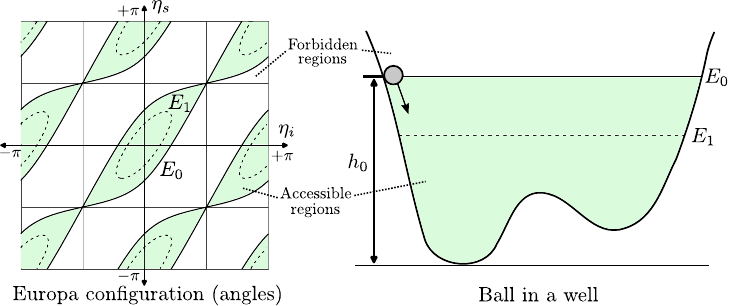}
\caption{Example zero-velocity curves and forbidden regions for a differentiated model of Europa and for a ball in a well. Both systems are bounded by the zero-velocity curves for some initial energy level $E_{0}$, which contracts to a new level $E_{1}$ with energy loss. The zero-velocity curves show the theoretical configuration bounds -- only the green part of the configuration space is accessible for a system with energy $E_{0}$.}
\label{fig:EnergiesConc}
\end{figure}

For a familiar analogy, consider the classical problem of a small ball placed at rest at height $h_{0}$ on the sloped side of a well. Here $E = \frac{1}{2}mv_{0}^{2} + V(h_{0}) = 0+ mgh_{0}$. Regardless of the shape of the well, if the ball returns to height $h_{0}$ it will have zero velocity due to conservation of energy, and at heights $h<h_{0}$ it will have velocity greater than zero, $v = \sqrt{2(\frac{E}{m} - gh)} = \sqrt{2g(h_{0} - h)}$. Heights $h>h_{0}$ are forbidden because they would imply imaginary velocity of the ball. Furthermore, with energy loss, the maximum height the ball can return to at time $t>0$ is simply $h(t) = E(t)/mg < E_{0}/mg$. The shape of the zero-velocity curves and the manner in which they contract allows for some useful insights into the spin-orbit coupling problem, particularly the instance of tidal locking. Figure~\ref{fig:EnergiesConc} conceptually depicts the zero-velocity curves for the differentiated model of Europa side-by-side with the simple ball-in-well analog.

In application of these equations to the general system Eq.~\eqref{Europa2Deq1}, note that generally we expect $C_{i} \gg C_{s}$. This can be easily shown by using the equations of the moments of inertia for a homogeneous sphere of radius $r_{i}$ and thin shell of radius $r_{s}$ and thickness $d$:
\begin{equation}
\label{Ci_moi}
C_{i} = \frac{2}{5}mr_{i}^{2} = \frac{8}{15}\pi r_{i}^{5}\rho_{i}
\end{equation}
\begin{equation}
\label{Cs_moi}
C_{s} = \frac{8}{3}\pi r_{s}^{4}d\rho_{s}
\end{equation}
\begin{equation}
\label{CiOCs}
\frac{C_{i}}{C_{s}} \sim \frac{1}{5}\frac{r_{i}^{5}}{r_{s}^{4}d}\frac{\rho_{i}}{\rho_{s}}
\end{equation}
Substituting for example $d \approx 24$ km, $r_{i} \approx 1440$ km, and $\rho_{i}/\rho_{s} \sim 3.2$, then $C_{i}/C_{s} \sim 28$. \cite{VanHoolst_Europa} gives $\rho_{s} \approx 920 \ \text{kg} \ \text{m}^{-3}$ and $\rho_{i}$ of at least $3000 \ \text{kg} \ \text{m}^{-3}$, and predict values of $C_{i}/C_{s}$ between 7 and 200. As a result of this imbalance, the motion of the interior tends to be less affected by the shell than vice-versa. 

We perform the same calculation for other icy bodies in the solar system differentiated into an icy surface, global water ocean, and solid interior. For some of these bodies, the thin-shell approximation is not appropriate, and $\gamma = C_{i}/C_{s}$ is given below:
\begin{equation}
    \label{GammaThickShell}
    \gamma = \frac{C_{i}}{C_{s}} = \left(\frac{\rho_{i}}{\rho_{s}}\right)\frac{r_{i}^{5}}{r_{s}^{5} - r_{0}^{5}}
\end{equation}
for shell (body) radius $r_{s}$ and ocean interface radius $r_{0} = r_{s} - d$. Note that this low-fidelity estimate neglects potential mantle-core differentiation of the solid interior. With this expression, we produce rough estimates of the values in Table~\ref{table:IcyMoonsGamma} for select ocean worlds in the solar system, as well as reasonable ranges, using recent references for densities and dimensions. The internal models of some of these worlds are quite uncertain at present, particularly Callisto and Ganymede.

\begin{table}[h!]
\centering
\caption{Solid Interior/Icy Shell Moment of Inertia Ratios for Presumed Ocean Worlds}
\label{table:IcyMoonsGamma}
\begin{tabular}{l|l|c}
 & Parameter estimates & Estimated $\gamma$ value \\ \hline 
Callisto  & $\rho_{i} = 3000$ kg/m\textsuperscript{3}, $\rho_{s} = 1000$ kg/m\textsuperscript{3}, & $\sim 13$                     \\
\ & $r_{i} = 2170$ km, $r_{s} = 2410.3$ km, $d = 70$ km & \ \\
Enceladus  & $\rho_{i} = 2400$ kg/m\textsuperscript{3}, $\rho_{s} = 925$ kg/m\textsuperscript{3}, & $1.6^{+0.6}_{-0.3}$                     \\
\ & $r_{i} = 190$ km, $r_{s} = 252.1$ km, $20<d<30$ km & \ \\
Europa    & $\rho_{i} = 3000$ kg/m\textsuperscript{3}, $\rho_{s} = 940$ kg/m\textsuperscript{3}, & $28.3^{+17.5}_{-13.3}$                     \\
\ & $r_{i} = 1440$ km, $r_{s} = 1560.8$ km, $d = 24.3^{+22.8}_{-1.5}$ km & \ \\
Ganymede  & $\rho_{i} \geq 3600$ kg/m\textsuperscript{3}, $\rho_{s} = 1000$ kg/m\textsuperscript{3}, & $\sim4$                     \\
\ & $r_{i} = 1800$ km, $r_{s} = 2631$ km, $d = 70$ km & \ \\
Titan  & $\rho_{i} = 2500$ kg/m\textsuperscript{3}, $\rho_{s} = 920$ kg/m\textsuperscript{3}, & $\sim7$                    \\
\ & $r_{i} = 2100$ km, $r_{s} = 2575$ km, $d = 80$ km & \                 
\end{tabular}
\end{table}
Our ranges of $\gamma$ for Europa and Enceladus are produced by exploring the bounds of constituent variables from literature. For Europa, we use the shell thickness and density ranges of \cite{Howell_2021}, along with seafloor radius estimates from \cite{Koh2022Europa}, and interior density ranges from \cite{Anderson1998} and \cite{VanHoolst_Europa}. Lower values of shell thickness produce higher limits of $\gamma$. To build the rest of Table~\ref{table:IcyMoonsGamma}, we borrow values from a range of other sources in literature. For Callisto, \cite{Journaux2020OceanWorlds, Spohn2003Europa}, for Enceladus, \cite{Cadek2019Enceladus, Hemingway2018Enceladus}, for Ganymede, \cite{Journaux2020OceanWorlds,Spohn2003Europa,Vance2014Ganymede}, and for Titan, \cite{Cadek2021,Journaux2020OceanWorlds}. The result we derive for Ganymede assumes a thin stable liquid region between the Ice VI layer and the silicate mantle. The presence of multiple intervening liquid layers between the high-pressure ice layers is a possibility \citep{Journaux2020OceanWorlds}. Our computed result of $\gamma$ for Ganymede would be higher if the silicate and metal interior is covered in the layers of high-pressure ice with no intervening liquid layers. We also assume an undifferentiated solid interior (no metallic core), resulting in a fairly high interior bulk density exceeding 3600 kg/m\textsuperscript{3} \citep{Vance2014Ganymede}. We derive the highest value of $\gamma$ for Europa out of all the most commonly speculated ocean worlds. We obtain the lowest value of $\gamma$ for tiny Enceladus, because the solid interior and ice shell have roughly comparable polar moments of inertia. 

\subsection{Models of Europa}
\subsubsection{The Van Hoolst Model}
\cite{VanHoolst_Europa} discusses the case that an icy satellite is differentiated into an ice shell and solid interior, nearly dynamically decoupled due to the existence of a deep ocean layer. The single-axis rotational equations of motion of the ice shell and interior are given below:
\begin{subequations}
\label{VH1}
\begin{align}
& C_{s}\ddot{\theta}_{s} + \frac{3}{2}\left( B_{s} - A_{s}\right) n^{2}\rho^{-3}\sin{\left( 2\left( \theta_{s} - f\right) \right)} = - K_{G}\sin{\left(2\left( \theta_{s} - \theta_{i}\right)\right)} 
\\
& C_{i}\ddot{\theta}_{i} + \frac{3}{2}\left( B_{i} - A_{i}\right) n^{2}\rho^{-3}\sin{\left( 2\left( \theta_{i} - f\right) \right)} =  K_{G}\sin{\left(2\left( \theta_{s} - \theta_{i}\right)\right)}
\end{align}
\end{subequations}
where $\rho = r/a$, $K_{G}$ is the constant of gravity-gradient coupling between the ice shell and interior, and $\theta_{s}$ and $\theta_{i}$ measure from from periapsis to the long axis of the ice shell and interior, respectively. The constant $K_{G}$ is given below assuming a solid interior:
\begin{equation}
\label{VH1b}
K_{G} = \frac{4\pi G}{5}\left(\rho_{s}\beta_{s} + (\rho_{0}-\rho_{s})\beta_{0}\right)\left(1 - \frac{\rho_{0}}{\rho_{i}}\right)\left(B_{i} - A_{i}\right)
\end{equation}
where $\rho_{s}$, $\rho_{0}$, and $\rho_{i}$ denote the densities of the shell, ocean, and interior, respectively, and similarly the $\beta$ terms are their respective equatorial flattenings. 

Eq. \eqref{VH1} makes some simplifying assumptions. First, only the longitudinal axis of rotation is considered. Also, this model assumes that the ice shell and interior are rigid -- neglecting the time-varying rotational effects on the polar moments of inertia $C_{i}$ and $C_{s}$, the zonal tidal effects on the polar flattening, and also dissipation within the shell for large differential angles $\theta_{s} - \theta_{i}$. Note that the time-varying rotational effects on a body's polar moment of inertia can be computed from the angular velocity $\dot{\theta}$ and the mean rotation rate $\omega$:
\begin{equation}
    \label{TVrot1}
    \frac{\delta C}{C} = \frac{5}{6}k_{2}\frac{\omega}{\pi G \rho}\left(\dot{\theta} - \omega\right)
\end{equation}
where $\rho$ is the mean density and $\delta C$ is the time-varying deviation induced in $C$. It can be shown that $\delta C/C \sim 10^{-4}$ for Europa, for reasonable values of $k_{2}$ and for the rotational rates considered in this work \citep{Wahr_EuropaTides, Sotin2009TidesAT}. Additionally, Eq.~\eqref{VH1} neglects any dynamic coupling effects from the subsurface ocean - including viscous torques and pressure gradient torques. At times it has been argued historically that for the purely longitudinal case, this coupling will be small in comparison to the gravity-gradient effects -- see e.g. \cite{VanHoolst_Europa} and \cite{Ojakangas1989EuropaPW}. 

While the effects of non-rigidity of the shell and interior could be important, and so could the effects of dynamic coupling due to the ocean, there are many interesting results and analyses available assuming a rigid shell and core that interact primarily gravitationally. 
The model given by Eq.~\eqref{VH1} was used in analysis in \cite{VanHoolst_Europa} to show that the existence and nature of a subsurface ocean and other aspects of Europa's interior can be studied via the librational response using the predictions of their model, which are modified from the classical rigid satellite case. However, there is more analysis that can be done with this model, beyond libration. Namely, full rotation can also be studied using the same equations.

The behavior of the averaged angles $\eta_{s} = \theta_{s} - nt$ and $\eta_{i} = \theta_{i} - nt$ is now studied. Making the necessary substitutions and neglecting the effects of orbit eccentricity, Eq.~\eqref{VH1} is transformed:
\begin{subequations}
\label{VH2}
\begin{align}
C_{s}\ddot{\eta}_{s} + \frac{3}{2}\left( B_{s} - A_{s}\right) n^{2}\sin{2\eta_{s}} + K_{G}\sin{\left(2\left( \eta_{s} - \eta_{i}\right)\right)} = \ & 0 \\
C_{i}\ddot{\eta}_{i} + \frac{3}{2}\left( B_{i} - A_{i}\right) n^{2}\sin{2\eta_{i}} - K_{G}\sin{\left(2\left( \eta_{s} - \eta_{i}\right)\right)} = \ & 0
\end{align}
\end{subequations}
This system of equations is that of two nonlinear pendula coupled by a nonlinear spring. It can be shown that this equation admits an energy integral, similar to the single-pendulum case:
\begin{equation}
\label{VH3}
E = \frac{1}{2}C_{i}\dot{\eta}_{i}^{2} + \frac{1}{2}C_{s}\dot{\eta}_{s}^{2} - \frac{3}{4}\left(B_{i} - A_{i}\right)n^{2}\cos{2\eta_{i}} - \frac{3}{4}\left(B_{s} - A_{s}\right)n^{2}\cos{2\eta_{s}} - \frac{1}{2}K_{G}\cos{\left(2\left(\eta_{s} - \eta_{i}\right)\right)}
\end{equation}
The last term is a coupling term and is a function of both coordinates $\eta_{s}$ and $\eta_{i}$. This type of term has previously been observed for other coupled oscillatory systems and can itself be useful for analysis \citep{DeSousa_CoupledSystemEnergy}. 

Similarly to before, the rate of change of the energy induced by dynamic tidal torque is given in terms of the separate orbit-averaged tidal torques on the interior and shell:
\begin{equation}
\label{VH4}
\dot{E} \approx \overline{L}_{\text{tidal},i}\dot{\eta}_{i} + \overline{L}_{\text{tidal},s}\dot{\eta}_{s}
\end{equation}
Due to the uncertainty in the nature of the non-gravitational torques at work in Europa, much of the work in this paper explores the tidal locking of Europa using the dynamical system previously introduced by Van Hoolst. However, Section 2.5 provides a thorough accounting of neglected torques.

\subsection{Application to Europa}
To apply our analysis to Europa, it is necessary to determine the range of possible scalings of the parameters in the dynamical models.
We primarily demonstrate this in-depth with the Van Hoolst model given by Eq.~\eqref{VH1}, but we generalize for a range of possible gravitational coupling strengths, which we show to be poorly constrained in the Van Hoolst model. There is additionally considerable uncertainty in the depth of Europa's ice shell, the depth of its ocean, and its internal composition. The historical thickness of the surface ice is also uncertain, and could have even had multiple freeze-thaw cycles as Europa's orbit has evolved over time. For our models, we neglect these unknowns and assume $\sim$24 km ice shell thickness and $\sim$100 km ocean depth, which is consistent with the range of values predicted in recent literature \citep{Anderson1998,Howell_2021,Wahr_EuropaTides}. With a mean radius of $r_{E} \approx 1560$ km, this yields $r_{i} \approx 1436$ km. In keeping with \cite{VanHoolst_Europa}, we use $\rho_{s} \sim 940 \ \text{kg} \ \text{m}^{-3}$ and $\rho_{i} \sim 3000 \ \text{kg} \ \text{m}^{-3}$. Next, an approximate scale of the gravitational coupling constant needs to be discerned. There is considerable uncertainty in the respective equatorial flattenings of the interior components of Europa, so the scale of the Van Hoolst model parameter $K_{G}$ is determined by a bulk flattening parameter $\tilde{\beta}$ defined below:
\begin{equation}
\label{Kg_scale}
\begin{split}
K_{G} = & \ \frac{4\pi G}{5}\rho_{0}\left(\frac{\rho_{s}}{\rho_{0}}\beta_{s} + (1-\frac{\rho_{s}}{\rho_{0}})\beta_{0}\right)\left(1 - \frac{\rho_{0}}{\rho_{i}}\right)\left(B_{i} - A_{i}\right) \\
= & \ \frac{4\pi G}{5}\rho_{0}\tilde{\beta}\left( B_{i} - A_{i}\right)
\end{split}
\end{equation}
Note $\rho_{0} \approx 1000 \ \text{kg} \ \text{m}^{-3}$ and $G = 6.674 \times 10^{-11} \text{m}^{3}/\text{kg} \ \text{s}^{2}$. For the purpose of analysis of the averaged system given in Eq.~\eqref{VH1}, it is more convenient to work with the following equations with non-dimensional time $\tau = nt$:
\begin{subequations}
\label{VH2_new}
\begin{align}
C_{s}\eta_{s}^{''} + \frac{3}{2}\left( B_{s} - A_{s}\right)\sin{2\eta_{s}} + \frac{K_{G}}{n^{2}}\sin{\left(2\left( \eta_{s} - \eta_{i}\right)\right)} = \ & 0 \\
C_{i}\eta_{i}^{''} + \frac{3}{2}\left( B_{i} - A_{i}\right)\sin{2\eta_{i}} - \frac{K_{G}}{n^{2}}\sin{\left(2\left( \eta_{s} - \eta_{i}\right)\right)} = \ & 0
\end{align}
\end{subequations}
For this system there exists modified energy integral $\hat{E}$:
\begin{equation}
\label{VH3_new}
\hat{E} = \frac{E}{n^{2}} = \frac{1}{2}C_{i}\eta_{i}^{'2} + \frac{1}{2}C_{s}\eta_{s}^{'2} - \frac{3}{4}\left(B_{i} - A_{i}\right)\cos{2\eta_{i}} - \frac{3}{4}\left(B_{s} - A_{s}\right)\cos{2\eta_{s}} - \frac{1}{2}\hat{K}_{G}\cos{\left(2\left(\eta_{s} - \eta_{i}\right)\right)}
\end{equation}
where $\hat{K}_{G} = K_{G}/n^{2}$ as below:
\begin{equation}
\label{khatG1}
\hat{K}_{G} = \frac{4\pi G}{5n^{2}}\rho_{0}\tilde{\beta}\left( B_{i} - A_{i}\right)
\end{equation}
Then $G\rho_{0} \approx 6.7 \times 10^{-8} \ \text{s}^{-2}$ and $n \approx 2\times 10^{-5} \ \text{s}^{-1}$, thus $\hat{K}_{G} \approx 398 \tilde{\beta} (B_{i} - A_{i})$. Noting that the equatorial flattening is $\beta = (r_{\text{max}} - r_{\text{min}})/r_{\text{max}}$, where $r_{\text{min}}$ and $r_{\text{max}}$ are measured along the equator, we expect values of the $\beta_{s}$ and $\beta_{i}$ to be on the order of $10^{-3}$ -- see e.g. \cite{Nimmo2007TheGS}. Because the other terms in $\tilde{\beta}$ are just ratios of densities, the bulk flattening parameter $\tilde{\beta}$ should also be on the same order. Conservatively, we can explore the problem for Europa with $\hat{K}_{G}$ at multiple orders in the following range: 
\begin{equation}
\label{khatG_range}
0.1(B_{i} - A_{i}) < \hat{K}_{G} < 10(B_{i} - A_{i})
\end{equation}
\subsubsection{Zero-Velocity Curves}
The curves of constant values of $E$ with $\eta^{'}_{s} = \eta^{'}_{i} = 0$ demarcate which parts of the parameter space are accessible at various energy levels. For the Van Hoolst model, there are two critical values of the energy that serve as the boundaries between qualitatively different behaviors:
\begin{equation}
\label{ZVCnew1}
\hat{E}_{\text{crit,1}} = \frac{E_{\text{crit,1}}}{n^{2}} = -\frac{3}{4}(B_{i} - A_{i}) + \frac{3}{4}(B_{s} - A_{s}) + \frac{1}{2}\hat{K}_{G}
\end{equation}
\begin{equation}
\label{ZVCnew2}
\hat{E}_{\text{crit,2}} = \frac{E_{\text{crit,2}}}{n^{2}} = \frac{3}{4}(B_{i} - A_{i}) + \frac{3}{4}(B_{s} - A_{s}) - \frac{1}{2}\hat{K}_{G}
\end{equation}
where the normalization scheme $\hat{E} = E/n^{2}$ is introduced for convenience to remove the appearance of the orbital mean motion $n$. The value $\hat{E}_{\text{crit,1}}$ is formed when the zero-velocity curves intersect at $\eta_{i} = \pm k\pi$, $k = 0, 1, 2 \ldots$, and $\eta_{s} = \pm l\pi/2$, $l = 1, 3, 5 \ldots$, and the value $\hat{E}_{\text{crit,2}}$ is formed for similar intersections at  $\eta_{i} = \pm k\pi/2$, $k = 1, 3, 5 \ldots$ and $\eta_{s} = \pm l\pi/2$, for nonzero odd integers $l$. There is a critical value of $\hat{K}_{G}$ for which $\hat{E}_{\text{crit,1}} = \hat{E}_{\text{crit,2}}$:
\begin{equation}
\label{Khat_G_crit}
\hat{K}_{G}^{*} = \frac{K_{G}^{*}}{n^{2}} = \frac{3}{2}\left(B_{i} - A_{i}\right)
\end{equation}
By careful study of the zero velocity curves, it can be shown that when $\hat{K}_{G} < \hat{K}_{G}^{*}$, it is possible for the body to decay into a state where the core is tidally locked while the ice shell continues to revolve independently. By contrast, if $\hat{K}_{G} > \hat{K}_{G}^{*}$, the final state will be one of synchronous rotation or simultaneous tidal locking of the ice shell and core. 

\begin{figure}[h!]
    \centering
    \subfloat[Indep. rotation possible, $\hat{K}_{G} = 0.5(B_{i}-A_{i})$]{\includegraphics[scale=1.0]{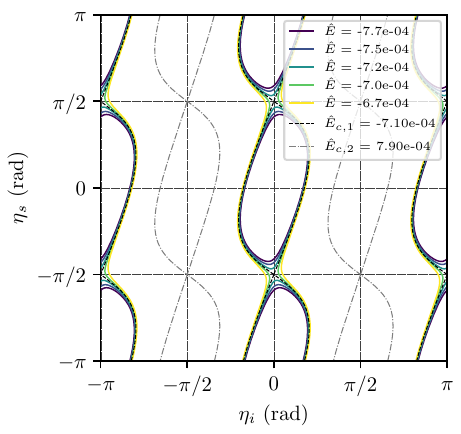}}
    \subfloat[Indep. rotation impossible, $\hat{K}_{G} =2.0(B_{i}-A_{i})$]{\includegraphics[scale=1.0]{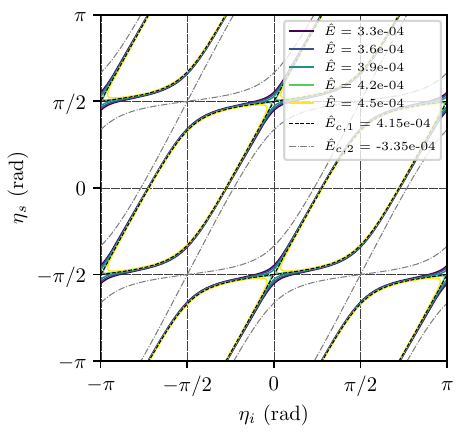}}
    \caption{Zero-velocity curves with constant values of $E/n^{2}$, where $C_{i} \equiv 1.0$. These plots illustrate the qualitatively different behavior of the zero velocity curved depending on whether $\hat{K}_{G}$ is less than or greater than the critical value $\hat{K}_{G}^{*}$.}
    \label{fig:ZVC_final}
\end{figure}

Zero-velocity contours for these two different cases are illustrated with numerical results in Figure \ref{fig:ZVC_final}. Note for this study that we set $C_{i} \equiv 1.0$ for simplicity and assume $(B_{i} - A_{i})/C_{i} = (B_{s} - A_{s})/C_{s} = 0.0015$, then assume $C_{i}/C_{s} = 28$. The choice of $C_{i} \equiv 1$ merely normalizes the dynamic equations, and does not impact the realism of the simulations. It is interesting to note that the value of $\hat{K}_{G}$ of Europa is not known with sufficient precision to determine if it is greater or less than $\hat{K}_{G}^{*}$.

For the weakly coupled case given in Figure~\ref{fig:ZVC_final}(a), it can be shown that if the total energy $E < E_{\text{crit,1}}$, full super-synchronous rotation of either the ice shell or core is impossible, and both parts will be in states of libration. If $E_{\text{crit,2}} > E > E_{\text{crit,1}}$, the ice shell is free to rotate, but the core is still tidally locked. The core is only free to rotate if $E > E_{\text{crit,2}}$, at which point both the ice shell and core will tend to be in super-synchronous rotation states. For the strongly coupled case given in Figure~\ref{fig:ZVC_final}(b), the ice shell and core will tend to co-rotate, with full rotation of both ice shell and core possible for $E > E_{\text{crit,2}}$. For $E_{\text{crit,1}} > E > E_{\text{crit,2}}$, full revolution of the ice shell requires full revolution of the core, and vice-versa. It is impossible for the core to be bounded while the ice shell continues to revolve, except for very high energy cases with $E > E_{\text{crit,1}}$. When $E > E_{\text{crit,1}}$, the zero-velocity curve boundaries shrink into inaccessible ``islands" of states that the system cannot occupy, and paths that permit unbounded motion in $\eta_{s}$ with bounded oscillations in $\eta_{i}$ become theoretically possible. 
Finally, when $E < E_{\text{crit,2}}$, both the ice shell and core become tidally locked. Note that it in some instances, the energy level will be sufficiently high to permit rotation of the ice shell and/or core, but all or part  of the system will be in a bounded rotation state. This is because the energy level only provides limits on \textit{admissible} behavior.

\begin{figure}[h!]
\centering
\includegraphics[]{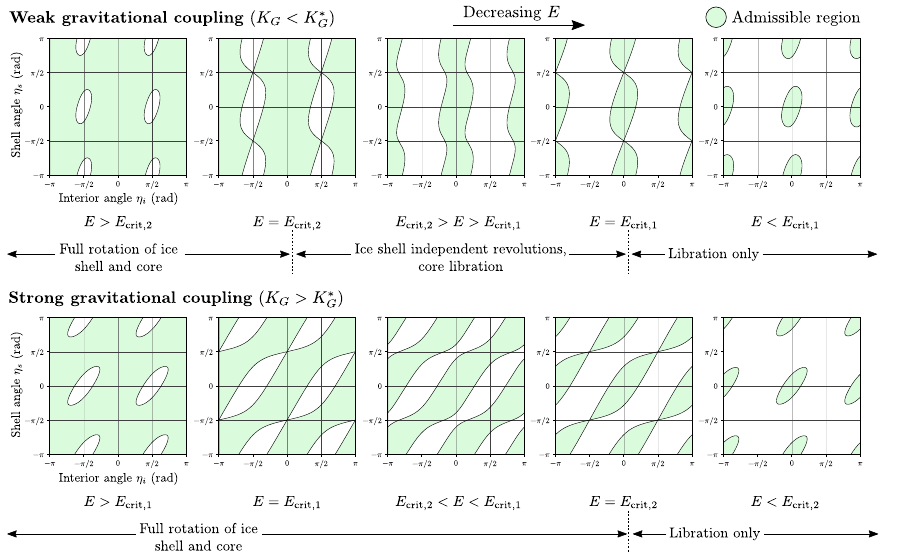}
\caption{Tidal locking and system energy using the Van Hoolst model with $\kappa_{i} = \kappa_{s}$, and $\gamma=C_{i}/C_{s} = 28$. As the total energy decreases over time due to tidal dissipation, the accessible configurations of ice shell and interior become increasingly limited, and the forbidden region grows. With weak gravitational coupling ($K_{G} < K_{G}^{*}$), the accessible region changes in a way that allows independent ice shell revolutions while the solid interior is tidally locked. With strong gravitational coupling ($K_{G} > K_{G}^{*}$), this is not permitted, and the ice shell and solid interior lock simultaneously.}
\label{fig:Energies}
\end{figure}

The two aforementioned cases show two very different possible dynamical evolutions of the rotating moon, depending on the strength of the gravity gradient torque between ice shell and core. 
In the case that $\hat{K}_{G} < \hat{K}_{G}^{*}$, energy loss from an initially super-synchronous rotation state can tidally lock the interior before the ice shell. However, if $\hat{K}_{G} > \hat{K}_{G}^{*}$, energy loss from an initially super-synchronous rotation state will eventually result in the simultaneous locking of the interior and ice shell. These cases are illustrated via numerical simulation later in the paper, and these conclusions are summarized graphically in Figure~\ref{fig:Energies}. Note that to get a holistic understanding of permitted motion for a particular case of interest, this figure should be reproduced with the desired value of $\gamma = C_{i}/C_{s}$ and the desired $\kappa_{i} =  (B_{i} - A_{i})/C_{i}$ and $\kappa_{s} =  (B_{s} - A_{s})/C_{s}$, because these will influence the exact shape of the zero-velocity curves.

\subsubsection{Simulations with the Averaged Dynamics}
\begin{figure}[h!]
    \centering
    \subfloat[Rotation Angles]{\includegraphics[scale=0.8]{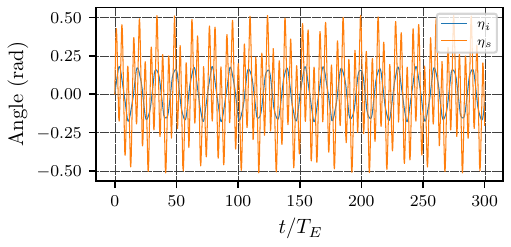}}
    \subfloat[State Bounded by Zero-Velocity Curves]{\includegraphics[scale=0.8]{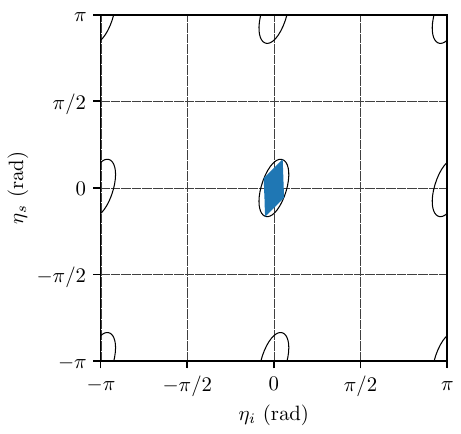}}
    \caption{Zero-velocity curves for a case with tidally locked interior and shell, and no energy dissipation. The zero-velocity curves demarcate the maximum extent of the state $\{\eta_{i}, \eta_{s} \}$.}
    \label{fig:ZVCfEx1}
\end{figure}

Numerical simulations of Eq.~\eqref{VH1} can be used to validate the fundamentals of the preceding energy analysis by plotting the behavior of the system in 2D with the relevant zero-velocity curves to illustrate how the state is bounded. Note for these simulations, the dynamic tidal torques are not considered, so the system conserves energy.
For the first example, consider the case of bounded librations of both ice shell and core, with the energy insufficient to permit full rotations. The relevant physical parameters and initial conditions are given in the first entry of Table \ref{table:SimSet1}.  The oscillations of the interior and the ice shell are given in Figure~\ref{fig:ZVCfEx1}(a). The states are also plotted in Figure~\ref{fig:ZVCfEx1}(b), which shows the evolving angle pair $\{\eta_{i}, \eta_{s} \}$ in blue, bounded externally by the black zero-velocity curve for energy $\hat{E} = f(\eta_{i,0}, \eta_{s,0}, \eta_{i,0}^{'}, \eta_{s,0}^{'})$. 

\begin{table}[h!]
\centering
\caption{Parameters for Simulations with Averaged Dynamics}
\label{table:SimSet1}
\begin{tabular}{ll|l}
Parameter                                                     & Simulation 1 & Simulation 2  \\ \cline{1-3}
\multicolumn{1}{l|}{Polar mom. inert., $C_{i}$}                & $C_{i} \equiv 1$             & $C_{i} \equiv 1$          \\
\multicolumn{1}{l|}{$\gamma = C_{i}/C_{s}$}                   & 28.0             & 28.0             \\
\multicolumn{1}{l|}{$\kappa_{j} = \frac{B_{j}-A_{j}}{C_{j}}$} &  $\kappa_{i} = \kappa_{s} = 0.0015$  & $\kappa_{i} = \kappa_{s} = 0.0015$              \\
\multicolumn{1}{l|}{ Grav. coupling, $\hat{K}_{G}$}  &  $0.5(B_{i}-A_{i})$ & $0.5(B_{i}-A_{i})$             \\
\multicolumn{1}{l|}{Initial orientations $(\eta_{i}, \eta_{s})$, deg}        &  $0,0$     & $-15, -70$              \\
\multicolumn{1}{l|}{Initial angular velocities $(\dot{\eta}_{i}, \dot{\eta}_{s})$, deg/s}          &  $10^{-5},10^{-4}$     & $-4\times 10^{-5}, 6.75 \times 10^{-5}$              \\
\multicolumn{1}{l|}{Duration (num. Europa orbits)}                             & 300             & 600           
\end{tabular}
\end{table}
\begin{figure}[h!]
    \centering
    \subfloat[Rotation Angles]{\includegraphics[scale=0.8]{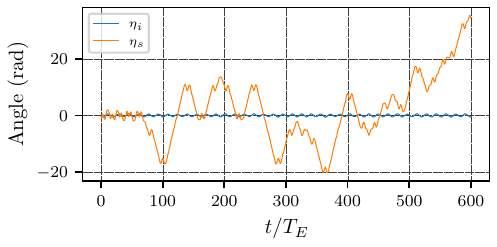}}
    \subfloat[State Bounded by Zero-Velocity Curves]{\includegraphics[scale=0.8]{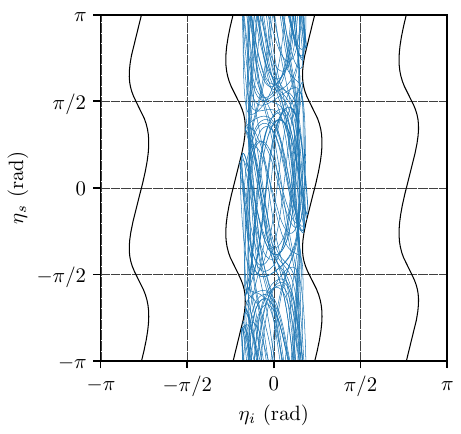}}
    \caption{Zero-velocity curves for a case with tidally locked interior, unbounded shell movement, and no energy dissipation. Zero-velocity curves demarcate the maximum extent of the state $\{\eta_{i}, \eta_{s} \}$.}
    \label{fig:ZVCfEx2}
\end{figure}
For the second example, given in the second entry of Table \ref{table:SimSet1}, the energy level and gravitational coupling constant $\hat{K}_{G}$ are chosen such that the interior is tidally locked but the ice shell is free to revolve.  The oscillations of the interior and the ice shell are given in Figure~\ref{fig:ZVCfEx2}(a), and these states are also plotted in Figure~\ref{fig:ZVCfEx2}(b), which shows the angles $\{\eta_{i}, \eta_{s} \}$ bounded externally by the zero-velocity curve for energy $\hat{E} = f(\eta_{i,0}, \eta_{s,0}, \eta_{i,0}^{'}, \eta_{s,0}^{'})$. Note that in this case the zero-velocity curves provide no constraints on the motion of $\eta_{s}$, but restrict the maximum oscillations of $\eta_{i}$, depending on the value of $\eta_{s}$. The behavior of the system for this case is quite complex, but the zero-velocity curves clearly show the fundamental limitations on the motion of the system. There is a simple underlying explanation for the chaotic shell motion when the orientation of the interior is also considered. Figure \ref{fig:DeltaHist} shows the differential angle $\delta = \eta_{s} - \eta_{i}$ between shell and core at the instant the ice shell reverses direction, with wrapping $|\delta| \leq \pi$. Large differential angles between the shell and interior generate a powerful torque on the ice shell, occasionally slowing the ice shell's differential rotation until it reverses direction. The torque is zero for $\delta = 0$, $\delta = \pm\pi/2$, and $\delta = \pm\pi$, explaining the drop-offs in turn-around instances near these values.
\begin{figure}[h!]
\centering
\includegraphics[]{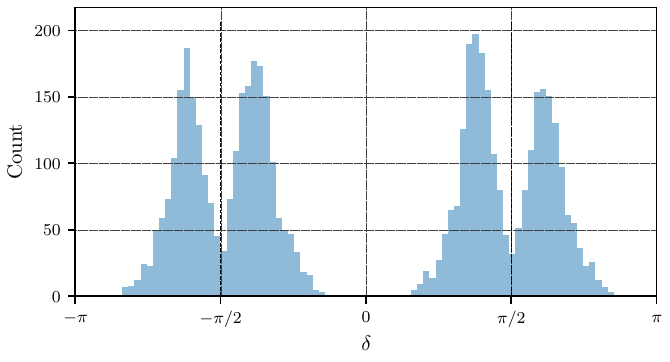}
\caption{Histogram of $\delta = \eta_{s} - \eta_{i}$, at the instant the ice shell reverses direction ($\eta_{s}^{'} = 0$), with wrapping $|\Delta| \leq \pi$. This shows that the relative orientation of the ice shell and interior strongly governs the behavior of the ice shell, and that the gravitational coupling between shell and core is responsible for the chaotic shell rotations. Generated by simulation of the chaotic solution (second entry of Table \ref{table:SimSet1}) for $5\times 10^{4}$ orbits.}
\label{fig:DeltaHist}
\end{figure}

\begin{figure}[h!]
\centering
\includegraphics[]{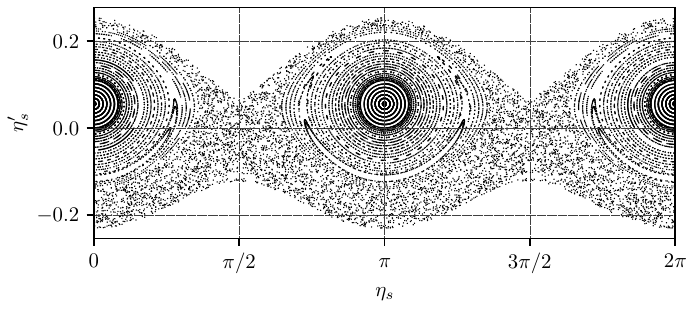}
\caption{Poincaré map for the case that the solid interior is restricted to librations, but the shell is potentially free to revolve, and gravitational coupling is weak. Shown are the rotation angle of the ice shell $\eta_{s}$ and the dimensionless rate $\eta_{s}^{'} = \frac{\text{d}}{\text{d}\tau}\left(\eta_{s}\right)$, where $\tau = nt$. The map shows a periodic solution when the shell and core librate in unison, surrounded by quasi-periodic solutions for independent and out-of-phase librations of shell and core. Beyond these regular libration solutions, which form closed curves in the map given sufficient time, the solution space is chaotic. The chaotic solutions are given as scattered and unorganized points outside the libration solution region.}
\label{fig:Poincare1}
\end{figure}

The complex behavior of the second example given in Figure~\ref{fig:ZVCfEx2} prompts a broader exploration of the possible behaviors exhibited by the system at an energy level $E_{\text{crit},2} > E > E_{\text{crit},1}$, where it is possible for the ice shell to undergo full revolutions independent of the tidally locked interior. Choosing an appropriate energy level $E = \frac{1}{2}\left(E_{\text{crit},1} + E_{\text{crit},2}\right)$ and using the same parameters from the Simulation 2 entry of Table \ref{table:SimSet1}, a Poincaré map is constructed, with the Poincaré surface defined as positive ($\eta_{i}^{'} > 0$) crossings of $\eta_{i} = 0$. The crossings are catalogued primarily from initial points with $\eta_{i}(t_{0}) = 0$, $\eta_{s}^{'}(t_{0}) = 0$, and $\eta_{s}(t_{0}) \in [0, 2\pi]$. The unspecified value of $\eta_{i}^{'}(t_{0})$ can be determined from the system energy. The resulting map is given in Figure~\ref{fig:Poincare1}. The map shows that chaotic motion of the ice shell constitutes a large part of the solution space for the system at this energy level. Whether or not Europa would undergo such chaotic behavior depends on the initial conditions and also the physical parameters governing the system - in particular the strength of gravitational coupling. It is noted that if the gravitational coupling is strengthened, the region of regular libration solutions expands, and the fraction of solutions exhibiting chaotic behavior is reduced. Additionally, the structure of the quasi-periodic solutions depends on the choice of values of $\kappa_{i}$ and $\kappa_{s}$. However, the same broad structure is generally observed: libration-based solutions near $\eta_{s} = 0$ and $\pi$, surrounded by a region of chaos.

\subsubsection{Simulations with Tidal Locking}
This section uses numerical simulations of the unaveraged and tidally forced dynamics to show, for initially super-synchronous rotation, how energy decay from the dynamic tidal torques can produce different final states, and to verify the conclusions of the energy analysis of this system. The dynamics are simulated with Europa's true anomaly $f$ as the independent variable, $\left( \ \right)^{'} = \frac{\text{d}}{\text{d}f}\left( \ \right)$, and the libration angles $\psi_{j} = \theta_{j} - f_{j}$ as the independent variables, using the following equations:
\begin{subequations}
\label{VHfdim}
\begin{align}
& \psi_{s}^{''} - 2\frac{r}{a}\frac{e\sin{f}}{1-e^{2}}\psi_{s}^{'} + \frac{3}{2}\left(\frac{B_{s} - A_{s}}{C_{s}}\right)\frac{r}{a(1-e^{2})}\sin{2\psi_{s}} = 2\frac{r}{a}\frac{e\sin{f}}{1-e^{2}} - \frac{K_{G}\sin{\left(2\left(\psi_{s} - \psi_{i}\right)\right)}}{C_{s}\dot{f}^{2}} + \frac{L_{\text{tidal},s}}{C_{s}\dot{f}^{2}}\\
& \psi_{i}^{''} - 2\frac{r}{a}\frac{e\sin{f}}{1-e^{2}}\psi_{i}^{'} + \frac{3}{2}\left(\frac{B_{i} - A_{i}}{C_{i}}\right)\frac{r}{a(1-e^{2})}\sin{2\psi_{i}} = 2\frac{r}{a}\frac{e\sin{f}}{1-e^{2}} + \frac{K_{G}\sin{\left(2\left(\psi_{s} - \psi_{i}\right)\right)}}{C_{i}\dot{f}^{2}} + \frac{L_{\text{tidal},i}}{C_{i}\dot{f}^{2}}
\end{align}
\end{subequations}
where $\dot{f} = h/r^{2}$, and the parameters $r, \ a, \ e$ are Europa's time-varying distance from Jupiter, constant semimajor axis, and eccentricity. Lastly, $L_{\text{tidal},j}$ is the dynamic tidal torque acting on body $j$. Using the simple MacDonald tidal model, it takes the following form:
\begin{equation}
\label{TidesEq1_2B} 
L_{\text{tidal},j} = - \tilde{D}_{j}\left(\frac{a}{r}\right)^{6}\text{sign}(\dot{\psi}_{j})
\end{equation}
where traditionally $\tilde{D}$ could be defined in terms of body parameters \citep{SSD_1999}:
\begin{equation}
\label{TidesEq2_2B}
\tilde{D} = \frac{3}{2}\frac{k_{2}}{Q_{s}}\frac{n^{4}}{G}R_{s}^{5}
\end{equation}
where $k_{2}$ is the tidal Love number, $Q$ is the tidal dissipation function, and $R_{s}$ is the satellite reference radius. The parameterization of $\tilde{D}$ as given by Eq. \eqref{TidesEq2_2B} render Eq. \eqref{TidesEq1_2B} as the MacDonald tidal torque, which assumes a constant lag/lead angle for the tidal bulge. For the simulations which follow, the approach is less granular. The term $\tilde{D}$ is generally chosen to be large enough for tidal locking to occur on reasonable simulation timescales, since the scale of the dynamic tidal torque influences the pace of the system energy loss -- see e.g. Eq. \eqref{VH4}. This is similar to the approach in \cite{Goldreich_TAJ1966}. Note that we do not the consider the complex long-timescale variations in the eccentricity and semimajor axis of Europa's perturbed orbit in this work. These are discussed in \cite{Bills_EuropaRotational} and references therein. 

\cite{EfroimskyTidal} discuss an alternate approach to Goldreich for the 1:1 spin-orbit resonance. They argue that the MacDonald tidal torque model in Goldreich and Peale's analysis is not rigorous because it conflates constant time lag with constant phase lag. Their following alternate expression serves as a corrected analog of the MacDonald tidal torque:
\begin{equation}
\label{EfrTidal1}
L_{\text{Tidal},j} = \frac{3}{2}GM_{J}^{2}k_{2}\frac{R^{5}}{r^{6}}\Delta t \cdot 2(\dot{f} - \dot{\theta}_{j}) + \mathcal{O}(i^{2}/Q) + \mathcal{O}(en\Delta t/Q)
\end{equation}
The term $\Delta t$ is the tidal time delay term -- in essence, because the tidal response of a satellite with real (not perfectly elastic) rheology is delayed, we can imagine that the tide raised on the satellite at time $t$ is the same as what would be instantaneously raised on the satellite by the tide-raising planet from its location at time $t - \Delta t$ -- an assumption at the heart of the derivation of the MacDonald tidal model. \cite{EfroimskyTidal} argue that the trouble with the MacDonald tidal approach is threefold. First, the assumption that all rheological modal contributions of the tidal response all experience the same time lag term $\Delta t$ is unphysical and restrictive. In reality, we would expect a Fourier-like response with distinct frequencies and lags summing to produce the necessary tidal potential and ensuing torque. Second, the MacDonald tidal approach derives the following relationship between a ``bulk" quality factor $Q$ and the resulting phase shift:
\begin{equation}
\label{MacDonaldsQ}
Q = \frac{1}{\sin{\text{``$\Delta$"}}} = \frac{1}{\sin{|2(\dot{\theta} - \dot{f})\Delta t|}}
\end{equation}
where $\Delta = 2\epsilon_{g}$ is the tidal phase angle, twice the geometric lag angle. The geometric lag angle is the angle between the long axis of the time-delayed tidal bulge and the planet-satellite line, the origin of the dynamic tidal torque, directly proportional to $\dot{\psi}$ with $\text{sgn}(L_{\text{Tidal}}) = -\text{sgn}(\dot{\psi})$. For the case of a bulk tidal response with time lag $\Delta t$, it is not clear that the ``$\Delta$" and Q of Eq.~\eqref{MacDonaldsQ} have the same physical interpretation as the classical unimodal phase shift and quality factor for a driven sinusoidal oscillator with a single frequency -- see e.g. \cite{SSD_1999}. This relationship should rather exist for each modal contribution to the total tidal response. Lastly, the geometric lag angle varies in time, so the resulting tidal torque should also vary in time, which is not captured by relations like Eq.~\eqref{TidesEq1_2B}. For an eccentric orbit, for short timescales (e.g. assuming constant $\dot{\theta}$), we can get a sense of the oscillations in $\epsilon_{g}$ as below for mean anomaly $M$ and mean motion $n$:
\begin{equation}
\label{epsilon_g}
\epsilon_{g} = \left(\dot{f} - \dot{\theta} \right)\Delta t \approx \left(n(1 + 2e\cos{M} + \frac{3}{2} e^{2}\cos{2M}) - \dot{\theta} \right)\Delta t
\end{equation}
Thus the oscillations in our $\epsilon_{g}$ term are simply $\mathcal{O}(e)$ over the course of an orbit, but can be much larger over the course of tidal despinning, for example, because there will be large changes in the value of $\dot{\theta}$.

For the corrected MacDonald-like approach, we still assume that $\Delta t$ is the same for all Fourier modes in the tidal response of the satellite, similarly to the classical treatment, but we restore the time-varying strength of the tidal torque using Eq.~\eqref{EfrTidal1}, as discussed in \cite{EfroimskyTidal}. We also need to establish comparable scaling of the modified and classical MacDonald torque expressions to facilitate one-to-one numerical comparisons. 
During de-spinning, we derive the following scaling via comparison of Eqs.~\eqref{TidesEq1_2B}-\eqref{TidesEq2_2B} and \eqref{EfrTidal1}:
\begin{equation}
\label{EfrTidal2}
k_{2}\cdot\Delta t \cdot 2(\dot{\theta} - \dot{f}) \sim \frac{k_{2}}{Q}
\end{equation}
where a direct functional equality should not be implied. Note that the value of $\dot{\theta}$ varies over the course of our simulations, which generally start in a slightly super-synchronous spin state, e.g. $2 > \dot{\theta}/n > 1$. To obtain a tidal torque scaling comparable to the classical MacDonald approach, we derive the following relationship between our specified $\tilde{D}$ and a resulting $k_{2}\Delta t$:
\begin{equation}
\label{EfrTidal3}
k_{2}\Delta t \sim \frac{1}{3}\frac{G}{n^{5}\Gamma}R_{s}^{-5}\tilde{D}
\end{equation}
with the unavoidable free parameter $\Gamma$ taking the place of $(\dot{\theta} - \dot{f})/n$ (see Eqs.~\eqref{EfrTidal1} and \eqref{TidesEq1_2B}), thus $1 > \Gamma > 0$. This is because the magnitude of the MacDonald tidal torque strength does not decrease with de-spinning, whereas the corrected expression does. In particular, $\Gamma$ sets the scale of the tidal strength such that the corrected tidal torque equals the classical MacDonald tidal torque when $\dot{\psi}_{j} = \Gamma n$. This can be seen by simplifying and rewriting Eq.~\eqref{EfrTidal1}:
\begin{equation}
\label{EfrTidalF}
L_{\text{tidal},j} = -\tilde{D}_{j}\left( \frac{a}{r}\right)^{6}\frac{1}{n\Gamma}\dot{\psi}_{j}
\end{equation}
Our procedure for choosing a suitable $\Gamma$ is to make it sufficiently small to achieve numerical settling (e.g. equilibrium energy level reached) over the desired timescale, which is typically a few thousand orbits, particularly for our large parameter studies where the run time is an important constraint. The numerical justification for this approach is discussed in Section 2.4.4, where we show that equilibrium energy is not strongly affected by the choice of $\Gamma$, analogously to the flexibility in choice of $\tilde{D}$ in the classical MacDonald tidal model.\footnote{See Goldreich's discussion about scaling the MacDonald tidal torque on p. 5 in \cite{Goldreich_TAJ1966}. Unlike Goldreich's plots (for the 1D problem), our energy plots (for a 2D problem) retain characteristic short-period oscillations in the energy parameter because we are not making any use of averaging in our numerical implementations. Also, as a historical note, Goldreich's single-run results were performed on a 1960s IBM 7094, for which run time for even a single simulation could be very long. By contrast, our parameter studies of $\sim$13,500 simulations each are performed on a modern computer cluster. Nonetheless, the problem of reasonable run time remains for us as well.} Note that we can easily identify equilibrium spin states via the energy integral, because we know $\overline{\dot{E}} = 0$ when equilibrium is achieved. There will still be non-secular variations in $E(t)$ because the numerical simulations are of higher fidelity than the model used to obtain the energy integral, which is essentially orbit-averaged, and also neglects dynamic tidal torque effects.

It has previously been noted that using the more complicated Darwin-Kaula tidal model, more reflective of realistic rheologies, the non-synchronous rotation state can disappear as an equilibrium solution altogether \citep{MakarovEfroimsky2013}. Given the context of our own previous finding that Europa seems to be in a tidally locked state at present \citep{Burnett_Icarus} and not in non-synchronous rotation as some have previously suggested \citep{HoppaScience1999,Geissler1998,GreenbergEuropaRotation2002,Greenberg2003TidalStress}, we briefly outline what is entailed in a use of the Darwin-Kaula torque for tidal locking studies of Europa. Application of the Darwin-Kaula torque involves computing the tidal potential with a particular lag for each tidal mode, e.g.:
\begin{subequations}
\label{DK1}
\begin{align}
& U_{\text{Tidal}} = U(\bm{r},\bm{r}^{*},\omega_{lmpq}, \Delta t_{lmpq})  \\
& L_{\text{Tidal}} = -M_{J}\frac{\partial U_{\text{Tidal}}}{\partial \theta}
\end{align}
\end{subequations}
Where $\bm{r}^{*}$ denotes the location of the tide-raising perturber, and the resulting potential is felt at $\bm{r}$. The $\omega_{lmpq}$ are the tidal modal frequencies, $\Delta t_{lmpq}$ are the corresponding modal time lags, and $\theta$ is the sidereal angle as before. See Eqs. (99) and (106) in \cite{Efroimsky-2012b} for the full expressions for Eq.~\eqref{DK1}, which are a complex series of functions of eccentricity and inclination. That work then considers a truncation of the series to $l = 2$ and $l = 3$, retaining terms to $\mathcal{O}(e^{2})$ and $\mathcal{O}(i^{2})$. Note that the MacDonald model is a unimodal model retaining only $l = m = 2$, the tidal potential is quadratic in the eccentricity and inclination functions, and furthermore that Europa's eccentricity and inclination are of comparable order i.e. $e = 0.009$ and $i = 0.0082$ radians. For an intermediary between the simple MacDonald model and the full Darwin-Kaula calculation, a truncated expansion of at most $\mathcal{O}(e)$ and $\mathcal{O}(i)$ would be quite valuable, if the dominant rheological modes are identified and only those are retained in the resulting truncated tidal potential and torque. We leave a more extensive numerical study of Europa tidal locking with such higher-fidelity rheological considerations to future work. With the ice shell and solid interior likely exhibiting dramatically different rheological responses, an accurate multimodal accounting of this response is an important issue with potentially major ramifications for our understanding of the dynamical evolution of the icy ocean worlds.

For simplicity and initial consistency with Goldreich and Peale, we start with results using the classical MacDonald tidal torque. Afterwards, we revisit some of these results with application of the corrected analog from \cite{EfroimskyTidal}. Two simulations highlight tidal locking of Europa for the two fundamental cases of interest -- (1) the case of weak interior gravity gradient torque coupling ($\hat{K}_{G} < \hat{K}_{G}^{*}$) and (2) the case of strong interior gravity gradient torque coupling ($\hat{K}_{G} > \hat{K}_{G}^{*}$). The data for these simulations is given in Table~\ref{table:EuropaDeSpin1}. The angles $\eta_{j}$ and angular rates $\dot{\eta}_{j}$ are used to characterize the initial conditions, but these can easily be converted via $\psi_{j} = \eta_{j} - f + nt$ and $\dot{\psi}_{j} = \dot{\eta}_{j} - \dot{f} + n$. Lastly, note $\psi_{j}^{'} = \frac{1}{\dot{f}}\dot{\psi}_{j}$. These same conversions are used to compute the value of the energy over the course of the simulations. The uncertain historical value of Europa's orbital eccentricity is a free variable in our simulations. 

\begin{table}[h!]
\centering
\caption{Parameters for Tidal Locking Simulations}
\label{table:EuropaDeSpin1}
\begin{tabular}{ll|l}
Parameter                                                     & Simulation 1 & Simulation 2  \\ \cline{1-3}
\multicolumn{1}{l|}{Polar mom. inert., $C_{i}$}                & $C_{i} \equiv 1$             & $C_{i} \equiv 1$          \\
\multicolumn{1}{l|}{$\gamma = C_{i}/C_{s}$}                   & 28.0             & 28.0             \\
\multicolumn{1}{l|}{$\kappa_{j} = \frac{B_{j}-A_{j}}{C_{j}}$} &  $\kappa_{i} = 0.0035$, $\kappa_{s} = 0.0015$  & $\kappa_{i} = 0.0035$, $\kappa_{s} = 0.0015$ \\
\multicolumn{1}{l|}{ Grav. coupling, $\hat{K}_{G}$}  &  $0.1(B_{i}-A_{i})$ & $2.5(B_{i}-A_{i})$             \\
\multicolumn{1}{l|}{ Dynamic tides, $\tilde{D}_{i}$, $\tilde{D}_{s}$}  & $\tilde{D}_{i} = 0.003K_{G}$, $\tilde{D}_{s} = 0.0015K_{G}$ & $\tilde{D}_{i} = 0.005K_{G}$, $\tilde{D}_{s} = 0.0005K_{G}$ \\
\multicolumn{1}{l|}{Initial orientations $(\eta_{i}, \eta_{s})$, deg}        &  $-80.76, -103.47$     & $0, 10.5$              \\
\multicolumn{1}{l|}{Initial ang. vel. $(\dot{\eta}_{i}, \dot{\eta}_{s})$, deg/s}          &  $2.72\times 10^{-5}, -1.31\times 10^{-5}$     & $1.65\times 10^{-4}, 1.65 \times 10^{-5}$              \\
\multicolumn{1}{l|}{$\hat{E}_{0}$, $\hat{E}_{\text{crit},1}$, $\hat{E}_{\text{crit},2}$}    & 0.0027, -0.0024, 0.0025   &  0.0031, 0.0018, -0.0017   \\
\multicolumn{1}{l|}{Europa orbit}                             & $a =  670900$ km, $e = 0.0094$, $f_{0} = 0$  & $f_{0} = 0$     \\
\multicolumn{1}{l|}{Duration, Europa orbits}                             & 900 (Fig. 9), 14000 (Fig. 10)    & 800           
\end{tabular}
\end{table}

For the first simulation, Europa is placed in an initially super-synchronous spin state, with the ice shell and core revolving in lockstep. After about 280 orbits, the interior tidally locks, and the ice shell transitions from super-synchronous spin to more erratic behavior, characterized by alternating prograde and retrograde rotations.
\begin{figure}[]
    \centering
    \subfloat[Rotation Angles]{\includegraphics[scale=1.0]{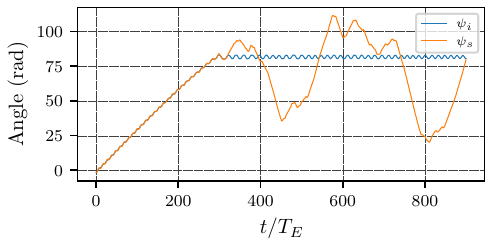}}
    \subfloat[Rotational Energy]{\includegraphics[scale=1.0]{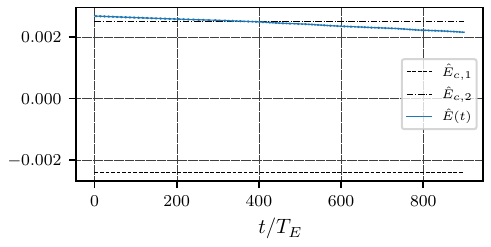}}
    \caption{Tidal locking of Europa's core, weak gravity-gradient coupling ($\hat{K}_{G} < \hat{K}_{G}^{*}$), with erratic ice shell movement. This short timescale demonstrates initially super-synchronous rotation, followed by the locking of Europa's core while its ice shell continues to revolve, with $\hat{E}(t) \gg \hat{E}_{\text{crit},1}$.}
    \label{fig:TidalDespinEx1}
\end{figure}
The erratic behavior of the ice shell continues until after 9000 orbits when it tidally locks as well. Figure~\ref{fig:TidalDespinEx1}(a) highlights the initially super-synchronous behavior and the irregular ice shell rotations after the interior tidally locks,  and Figure~\ref{fig:TidalDespinEx1}(b) shows that the locking shortly precedes the transition of $\hat{E}(t)$ to below the critical value for core tidal locking, $\hat{E}_{\text{crit},2}$. The full simulation is shown in Figure~\ref{fig:TidalDespinEx1b}. The tidal locking of the ice shell occurs just before the transition of $\hat{E}(t)$ to below the critical value for ice shell tidal locking, $\hat{E}_{\text{crit},1}$. This behavior is expected from the earlier analysis based on the zero-velocity curves in Section 2.4.1. Note that after the ice shell tidally locks, the energy decay rate slows significantly. Also, note that the ice shell rotations do not decouple from the interior for all cases with $\hat{K}_{G} < \hat{K}_{G}^{*}$. However, this behavior is quite common, depending on the choices of parameters and the initial conditions of the system. By contrast, in the case of $\hat{K}_{G} > \hat{K}_{G}^{*}$, it cannot occur at all.

\begin{figure}[]
    \centering
    \subfloat[Rotation Angles]{\includegraphics[scale=1.0]{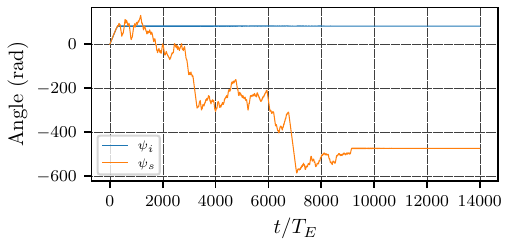}}
    \subfloat[Rotational Energy]{\includegraphics[scale=1.0]{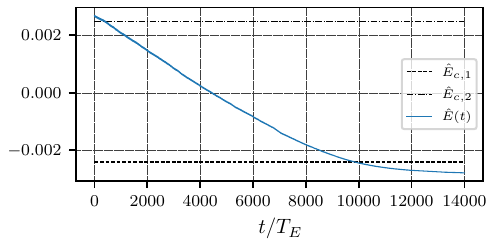}}
    \caption{Tidal locking of Europa's ice shell and core, weak gravity-gradient coupling ($\hat{K}_{G} < \hat{K}_{G}^{*}$). On this long timescale of 14000 Europa orbits ($\sim 136$ yrs), both the core and ice shell tidally lock. These locking instances occur shortly before $E(t)$ decays to the corresponding critical values for locking -- $\hat{E}_{\text{crit},2}$ for the core, and $\hat{E}_{\text{crit},1}$ for the shell.}
    \label{fig:TidalDespinEx1b}
\end{figure}
\begin{figure}[]
    \centering
    \subfloat[Rotation Angles]{\includegraphics[scale=1.0]{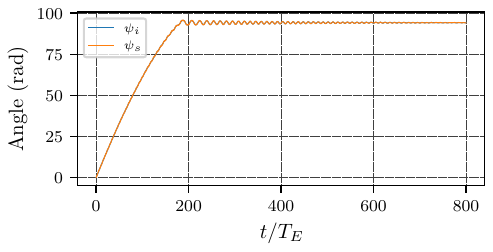}}
    \subfloat[Rotational Energy]{\includegraphics[scale=1.0]{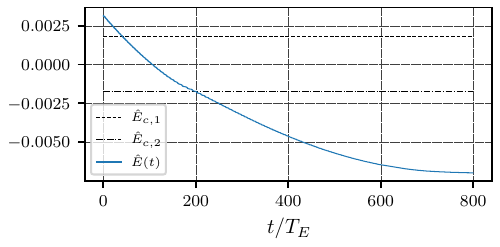}}
    \caption{Tidal locking of Europa's ice shell and core, strong gravity-gradient coupling ($\hat{K}_{G} > \hat{K}_{G}^{*}$). Over 200 Europa orbits, initially super-synchronous rotation of core and shell is tidally locked. Afterwards, librations are steadily reduced. The simultaneous locking occurs near when $E(t)$ decays to the corresponding critical value for simultaneous locking, $\hat{E}_{\text{crit},2}$.}
    \label{fig:TidalDespinEx2}
\end{figure}
For the second simulation, Europa is again placed in initially super-synchronous rotation with the core and ice shell revolving in unison. However, for this simulation, $\hat{K}_{G} > \hat{K}_{G}^{*}$, and numerical experiments for this case show that the ice shell and core generally tidally lock simultaneously. This is in contrast to the decoupling cases observed with $\hat{K}_{G} < \hat{K}_{G}^{*}$, and in line with the predictions from the earlier analysis with the zero-velocity curves in Section 2.4.1. The tidal locking in the example shown in Figure~\ref{fig:TidalDespinEx2} occurs after about 200 orbits, and coincides with the transition of $\hat{E}(t)$ to the critical value below which super-synchronous rotation is impossible -- $\hat{E}_{\text{crit},2}$.

\subsubsection{Non-synchronous Rotation}
An example with non-synchronous rotation as a final spin state is also provided, to illustrate the variety of final behaviors that are possible with a differentiated Europa. For this case, the simulation parameters are given in Table \ref{table:EuropaNSR1}. Note that Europa's eccentricity is increased to $0.05$ for this simulation. The scale of the dynamic tidal torques is chosen to allow for a reasonable simulation time to reach a final equilibrium state. With this choice of eccentricity, the specified asymmetry in the ice shell is sufficient to expect tidal locking, using Goldreich and Peale's classical result. In particular, the value of $\kappa_{s}= \frac{B_{s} - A_{s}}{C_{s}} = 0.0005$ exceeds the classical limit from Goldreich and Peale above which tidal locking is expected: Eq. ~\eqref{Ean2} yields a cutoff of $\kappa_{\text{max}} = \frac{B - A}{C} = 0.000464$. However, while $\kappa_{s}$ exceeds this value, $\kappa_{i}$ does not. 

\begin{table}[h!]
\centering
\caption{Parameters for Non-Synchronous Rotation Simulation 1}
\label{table:EuropaNSR1}
\begin{tabular}{ll|l}
Parameter                                                     & Value & \\ \cline{1-3}
\multicolumn{1}{l|}{Polar mom. inert., $C_{i}$}                & $C_{i} \equiv 1$             &    \\
\multicolumn{1}{l|}{$\gamma = C_{i}/C_{s}$}                   & 28.0             &  \\
\multicolumn{1}{l|}{$\kappa_{j} = \frac{B_{j}-A_{j}}{C_{j}}$} &  $\kappa_{i} = 0.0003$, $\kappa_{s} = 0.0006$  &  \\
\multicolumn{1}{l|}{ Grav. coupling, $\hat{K}_{G}$}  &  $0.6(B_{i}-A_{i})$ &  \\
\multicolumn{1}{l|}{ Dynamic tides, $\tilde{D}_{i}$, $\tilde{D}_{s}$}  & $\tilde{D}_{i} = 0.2K_{G}$, $\tilde{D}_{s} = 0.05K_{G}$ &  \\
\multicolumn{1}{l|}{Initial orientations $(\eta_{i}, \eta_{s})$, deg}        &  $0.0, 0.0$     &  \\
\multicolumn{1}{l|}{Initial ang. vel. $(\dot{\eta}_{i}, \dot{\eta}_{s})$, deg/s}          &  $4.17\times 10^{-5}, 8.34\times 10^{-5}$     &        \\
\multicolumn{1}{l|}{$\hat{E}_{0}$, $\hat{E}_{\text{crit},1}$, $\hat{E}_{\text{crit},2} \ \times 10^{-3}$}    & 0.3922, -0.1216, 0.1484   &    \\
\multicolumn{1}{l|}{Europa orbit}                             & $a =  670900$ km, $e = 0.05$, $f_{0} = 0$  &  \\
\multicolumn{1}{l|}{Duration, Europa orbits}                             & 5000  &     
\end{tabular}
\end{table}

\begin{figure}[htb!]
    \centering
    \subfloat[Rotation Angles]{\includegraphics[scale=1.0]{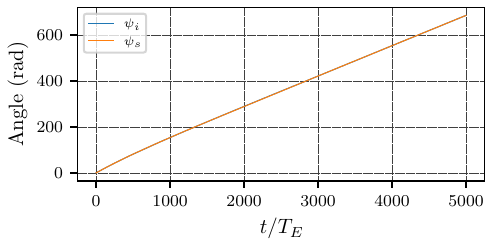}}
    \subfloat[Rotational Energy]{\includegraphics[scale=1.0]{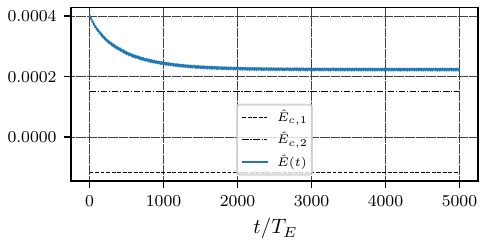}}
    \caption{Non-synchronous rotation of Europa's ice shell and core, weak gravity-gradient coupling ($\hat{K}_{G} < \hat{K}_{G}^{*}$).The energy decreases to a final equilibrium level $E_{f} > E_{c,2}$ and thus Europa does not settle into a tidally locked state.} 
    \label{fig:NSREx1}
\end{figure}

A numerical exploration of this case yields the results in Figure \ref{fig:NSREx1}. Europa does not settle into a tidally locked state in this simulation, despite the large mass asymmetry in the shell. The solid interior lacks sufficient asymmetry, and the ice shell is gravitationally dragged along into a slightly spun-up final non-synchronous rotation state, with super-synchronous final equilibrium energy $E_{f} > E_{c,2}$. This is despite the potential for energy dissipation via the dynamic tides. 

\begin{figure}[h!]
\centering
\includegraphics[]{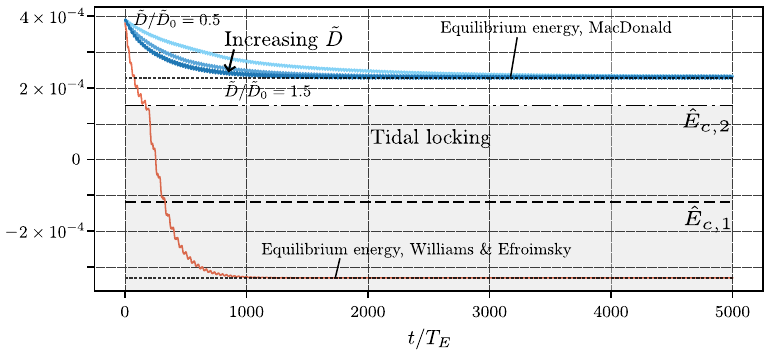}
\caption{Normalized energy, $E/n^{2}$, vs. time (Europa orbits). A non-synchronous rotation result is obtained with the classical MacDonald tidal torque \citep{MacDonaldTidal, SSD_1999}, but tidal locking is obtained with the modified tidal torque expression \citep{EfroimskyTidal} given by Eqs.~\eqref{EfrTidal1} and \eqref{EfrTidalF}. This figure starts with the same initial condition and generates the resulting numerical behavior with three scales of $\tilde{D}$ for MacDonald's model (three blue curves) in contrast with a lone curve for the Williams \& Efroimsky model. We find that the equilibrium spin state is not greatly affected by $\tilde{D}$, but the settling time is. This provides numerical justification for the practice of scaling up the MacDonald tidal torque to achieve more rapid numerical settling.}
\label{fig:MacDonaldSettling}
\end{figure}

Implementing the modified tidal torque in Eq.~\eqref{EfrTidal1} -- or equivalently, Eq.~\eqref{EfrTidalF}, we find that the equilibrium spin states for a given set of parameters are greatly modified from the results with the classical MacDonald tidal torque. In particular, consider the case parameterized in Table \ref{table:EuropaNSR1}, but now implemented with the modified tidal torque, Eq.~\eqref{EfrTidal1}, or equivalently \eqref{EfrTidalF}, with $\Gamma = 0.04$. This yields Figure~\ref{fig:MacDonaldSettling}. While the MacDonald tidal torque results in NSR for this case regardless of the value of $\tilde{D}$, the modified expression results in eventual tidal locking regardless of the value of $\Gamma$. In general, we find that the modified expression given by \cite{EfroimskyTidal} results in the boundary between NSR and tidal locking being shifted such that a significantly lower permanent mass asymmetry is capable of initiating tidal locking. 

\begin{figure}[h!]
\centering
\includegraphics[]{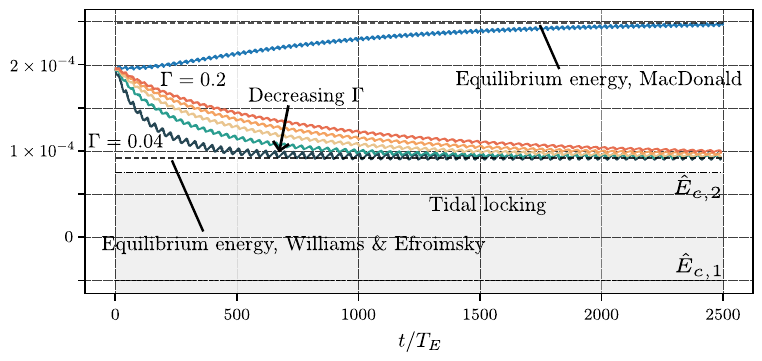}
\caption{Normalized energy, $E/n^{2}$, vs. time (Europa orbits). A non-synchronous rotation result is obtained with both the classical MacDonald tidal torque \citep{MacDonaldTidal,SSD_1999} and the modified tidal torque expression \citep{EfroimskyTidal} given by Eqs.~\eqref{EfrTidal1} and \eqref{EfrTidalF}. This figure starts with the same initial condition and generates the resulting numerical behavior with MacDonald's model (lone blue curve) in contrast with various settling times for the Williams \& Efroimsky model (colored curves) ranging uniformly from $\Gamma = 0.04$ to $\Gamma = 0.2$. We find that the equilibrium spin state is not greatly affected by $\Gamma$, but the settling time is. The Williams \& Efroimsky model is more dissipative than the MacDonald model, resulting in lower equilibrium energy, a relatively lower equilibrium spin, and generally a higher likelihood for tidal locking in direct numerical comparisons.}
\label{fig:EfroimskySettling}
\end{figure}

We can still obtain NSR as an equilibrium spin state in numerical simulations with the modified tidal torque, but we must greatly reduce the permanent moment of inertia asymmetry. Figure~\ref{fig:EfroimskySettling} shows the result of the case in Table \ref{table:EuropaNSRefr}, which gives a sufficiently low permanent asymmetry to induce NSR, at less than 1/3 of the value required by the classical MacDonald model, which is denoted by $\kappa^{*}$. For comparison we show a single result with the MacDonald model, for which the equilibrium spin state is actually faster than the initial condition. Note the corresponding equilibrium energy gap between the two types of solutions. To generate the results with the modified tidal torque, we vary the value of $\Gamma$ to manipulate the settling time, and note that this does not greatly influence the equilibrium energy level. A similar result was previously shown for NSR cases using a variable $\tilde{D}$ with the MacDonald tidal torque. Finally, note that the ice shell asymmetry for this simulation is rather high, but is dragged nonetheless into NSR via gravitational coupling with the super-synchronously rotating solid interior.

\begin{table}[h!]
\centering
\caption{Parameters for Non-Synchronous Rotation Simulation 2}
\label{table:EuropaNSRefr}
\begin{tabular}{ll|l}
Parameter                                                     & Value & \\ \cline{1-3}
\multicolumn{1}{l|}{Polar mom. inert., $C_{i}$}                & $C_{i} \equiv 1$             &    \\
\multicolumn{1}{l|}{$\gamma = C_{i}/C_{s}$}                   & 28.0             &  \\
\multicolumn{1}{l|}{$\kappa_{j} = \frac{B_{j}-A_{j}}{C_{j}}$} &  $\kappa_{i} = 0.3\kappa^{*}$, $\kappa_{s} = \kappa^{*}$  &  \\
\multicolumn{1}{l|}{ Grav. coupling, $\hat{K}_{G}$}  &  $0.6(B_{i}-A_{i})$ &  \\
\multicolumn{1}{l|}{ Dynamic tides, $\tilde{D}_{i}$, $\tilde{D}_{s}$}  & $\tilde{D}_{i} = 0.3K_{G}$, $\tilde{D}_{s} = 0.075K_{G}$ &  \\
\multicolumn{1}{l|}{ Range of $\Gamma$}  & $\Gamma \in [0.04, 0.2]$ & \\
\multicolumn{1}{l|}{Initial orientations $(\eta_{i}, \eta_{s})$, deg}        &  $0.0, 0.0$     &  \\
\multicolumn{1}{l|}{Initial ang. vel. $(\dot{\eta}_{i}, \dot{\eta}_{s})$, deg/s}          &  $2.92\times 10^{-5}, 5.83\times 10^{-5}$     &        \\
\multicolumn{1}{l|}{$\hat{E}_{0}$, $\hat{E}_{\text{crit},1}$, $\hat{E}_{\text{crit},2} \ \times 10^{-3}$}    & 0.1945, -0.0502, 0.0751   &    \\
\multicolumn{1}{l|}{Europa orbit}                             & $a =  670900$ km, $e = 0.05$, $f_{0} = 0$  &  \\
\multicolumn{1}{l|}{Duration, Europa orbits}                             & 2500  &     
\end{tabular}
\end{table}

\subsubsection{Tidal Locking and Ice Shell Asymmetry}

We now explore the end behavior of Europa (tidally locked or non-synchronous rotation) in several large 2D parametric studies with variables $\kappa_{i} = (B_{i} - A_{i})/C_{i}$ and $\kappa_{s} = (B_{s} - A_{s})/C_{s}$. The dynamic tidal torques are made large enough that equilibrium spin states can be achieved in feasible computational time, similar to the approach in Goldreich \& Peale. Note that an equilibrium spin state is one for which secular changes in system energy have stopped. The parameters used for the simulations are given in Table~\ref{table:TLPS1}.

\begin{table}[h!]
\centering
\caption{Simulation Parameters for Tidal Locking Parameter Studies}
\label{table:TLPS1}
\begin{tabular}{ll|l}
Parameter                                                     & Values & \\ \cline{1-3}
\multicolumn{1}{l|}{Polar mom. inert., $C_{i}$}                & $C_{i} \equiv 1$             &    \\
\multicolumn{1}{l|}{$\gamma = C_{i}/C_{s}$}                   & 28.0            &  \\
\multicolumn{1}{l|}{ Range of $\kappa_{i}$, $\kappa_{s}$}  &  $0.6 < \frac{\kappa_{i}}{\kappa^{*}} < 1.2$, $0.5 < \frac{\kappa_{s}}{\kappa^{*}} < 6.0$ &  \\
\multicolumn{1}{l|}{ Grav. coupling, $\hat{K}_{G}$}  &  $0.6(B_{i}-A_{i})$ or $1.6(B_{i}-A_{i})$ &  \\
\multicolumn{1}{l|}{ Dynamic tides, $\tilde{D}_{i}$, $\tilde{D}_{s}$}  & $\tilde{D}_{i} = 0.3K_{G}$, $\tilde{D}_{s} = 0.075K_{G}$ &  \\
\multicolumn{1}{l|}{Initial orientations $(\eta_{i}, \eta_{s})$, deg}        &  $0.0, 0.0$     &  \\
\multicolumn{1}{l|}{Initial ang. vel. $(\dot{\eta}_{i}, \dot{\eta}_{s})$, deg/s}          &  Variable     &        \\
\multicolumn{1}{l|}{Europa orbit}                             & $a =  670900$ km, $e = 0.05$, $f_{0} = 0$  &  \\
\multicolumn{1}{l|}{Duration, Europa orbits}                             & 2500  &     
\end{tabular}
\end{table}

For these studies, for each point, the minimum interior angular velocity for unbounded revolution is computed, then scaled by a constant factor to generate a super-synchronous interior rotation state. Similarly, the ice shell rotation state is scaled to ensure super-synchronous rotation of the ice shell. Values are chosen such that the ice shell and core generally revolve in lockstep in the beginning of the simulation. Starting with the case of $\gamma = 28$ and $\hat{K}_{G} = 0.6(B_{i}-A_{i})$, with the initial conditions as specified and with gravitational coupling $K_{G} < K_{G}^{*}$, Europa is initially in the high-energy case as depicted in the top left of Figure~\ref{fig:Energies}.

\begin{figure}[h!]
\centering
\includegraphics[]{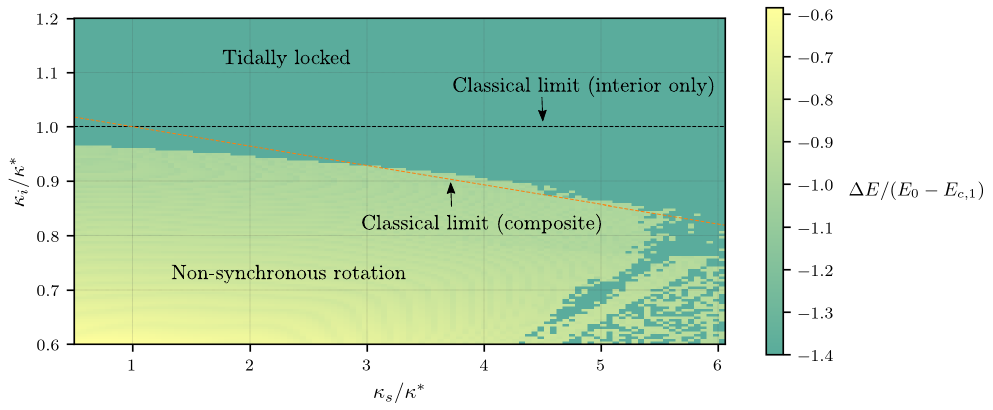}
\caption{Europa's equilibrium spin state vs. moment of inertia asymmetries $\kappa_{i}/\kappa^{*}$ and $\kappa_{s}/\kappa^{*}$, classical MacDonald torque, weak gravitational coupling case with $\hat{K}_{G}=0.6(B_{i}-A_{i})$ and $\gamma=28$. The value $\kappa^{*} = \frac{2}{3}\left(\frac{9.5\pi e^{2}}{2\sqrt{2}}\right)^{2}$ is the tidal locking $(B-A)/C$ limit from \cite{Goldreich_TAJ1966}, obtained using the classical MacDonald tidal torque. In the figure, $\Delta E = E_{f} - E_{0}$ is the energy decrease over 2500 orbits, and the dark cyan cases with $\Delta E/(E_{0}-E_{c,1}) < -1$ are all tidally locked. The classical tidal locking limit does not apply to Europa, and large ice shell asymmetry enables tidal locking for values of $\kappa_{i} < \kappa^{*}$. This result is obtained with the classical torque of \cite{MacDonaldTidal}. Over the same range of parameters, the modified tidal torque given by Williams \& Efroimsky results in universal tidal locking for this study.}
\label{fig:TLParam}
\end{figure}

The results of this first numerical study are given graphically in Figure~\ref{fig:TLParam}. This figure shows the drop in system energy over the simulation time span, with a clear boundary between tidally locked cases (for which $\Delta E/(E_{0}-E_{c,1}) < -1$) and non-synchronous rotation cases (with $\Delta E/(E_{0}-E_{c,1}) \gg -1$). The former are indicated with a dark cyan and the latter with a light green. Furthermore, within the non-synchronous rotation region, the coloration of an individual cell reflects the level of energy loss. The cases with least energy loss are more yellow, and these represent faster equilibrium non-synchronous rotation rates. 

By the choice of initial conditions, the moon cannot lose much energy before entering in a state of core or combined ice shell and core locking. Note that once $E<E_{\text{crit},2}$, tidal locking becomes inevitable -- equilibrium spin states with energies between $E_{\text{crit},2}$ and $E_{\text{crit},1}$ are absent from these numerical results. Overall it is observed that applying the classical condition given by Eq.~\eqref{Ean2} to the solid interior only (whose moments of inertia dwarf those of the shell) only roughly approximates the boundary between tidal locking and non-synchronous rotation. Additionally, as the moment of inertia asymmetry of the ice shell is increased, it becomes more likely for Europa to end up tidally locked. The boundary between tidally locked and NSR solutions migrates to lower values of $\kappa_{i}$ as $\kappa_{s}$ is increased. 

To better approximate the transition boundary between non-synchronous and tidally locked final spin states, consider application of the classic condition given by Eq.~\eqref{Ean2} to a differentiated but rigid Europa, where the shell is artificially geometrically fixed to the solid interior. With this assumption that the shell and interior are completely in lockstep, and noting that moments of inertia are additive, we obtain a linear boundary between tidal locking and NSR, $\kappa_{i}^{*} = \kappa^{*}\left(1 + \frac{1}{\gamma} - \frac{1}{\gamma}\frac{{\kappa}_{s}}{\kappa^{*}}\right)$, given as an orange line in Figure~\ref{fig:TLParam}. This linear assumption fails to accurately fit the nonlinear transition boundary, but serves as a rough approximation. The failure of this line to fit the transition boundary in the region with lower values of $\kappa_{s}/\kappa^{*}$ is notable -- the ability of the shell to move independently of the interior turns out to be dynamically important in the end behavior of the system. A shell with a permanent mass asymmetry that is classically thought too small to induce tidal locking is in fact able to induce tidal locking in these nonlinear simulations. We observe however that the linear slope fits the nonlinear simulation results well in the range $3 < \kappa_{s}/\kappa^{*} < 5.5$, as long as $\kappa_{i}$ is high enough to avoid the complex tidal locking behavior exhibited in the lower right corner of the plot.

The region in the lower right corner of Figure~\ref{fig:TLParam}, within the range $4 < \frac{\kappa_{s}}{\kappa^{*}} < 6$ and $0.6 < \frac{\kappa_{i}}{\kappa^{*}} < 0.85$ displays more complexity than the rest of the parameter space. In this region, independent rotation (and sometimes early intermittent locking) of the ice shell plays an especially important role by amplifying the tidal dissipation rate and driving Europa towards a fully tidally locked state. In this region, the core has much less permanent mass asymmetry than is expected to drive tidal locking, but the shell has a very large permanent mass asymmetry. The result is a complex variation of outcomes depending on the exact relative values of $\kappa_{s}/\kappa^{*}$ and $\kappa_{i}/\kappa^{*}$. Notable is the ``stable tongue" of tidally locked behavior extending deep into the region that is otherwise dominated by NSR. This feature bears some casual resemblance to the stability diagrams featured in classical and contemporary studies of systems which exhibit parametric resonance. See for example Figure 4 in \cite{KovacicEA_ParamResonance}. To extend the analogy, consider the simple damped Mathieu oscillator below with stiffness, damping, and perturbative parameters $\delta > 0$, $c$, and $\varepsilon$, for which $|\epsilon/\delta| \ll 1$. 
\begin{equation}
\label{Mathieu1}
\ddot{x} + c\dot{x} + (\delta + \varepsilon \cos{t})x = 0
\end{equation}
For the case of dissipative damping, in the absence of parametric forcing ($\varepsilon = 0$) this oscillator is stable, i.e. $x(t) \rightarrow 0$ for $\delta > 0$, and unstable for $\delta < 0$. However, when $\varepsilon > 0$, there are some combinations of $\delta$ and $\varepsilon$ which can stabilize the system even when $\delta < 0$. In our numerical study of Europa, we have a more complicated system -- two coupled second-order differential equations with additional forcing terms -- but we still have time-varying coefficients in the dynamics (including time-varying ``stiffness" terms), and we observe a region of mostly unsettled (i.e. non-synchronous) behavior for which there is settling (i.e. tidal locking) for a narrow range of parameters.

\begin{figure}[htb!]
    \centering
    \subfloat[Non-synchronous rotation case]{\includegraphics[scale=0.95]{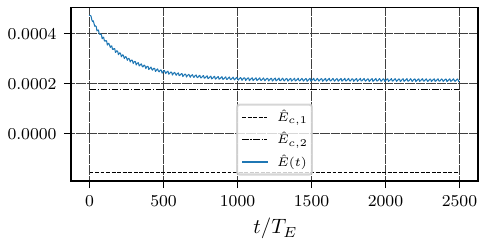}}
    \subfloat[Tidal locking case]{\includegraphics[scale=0.95]{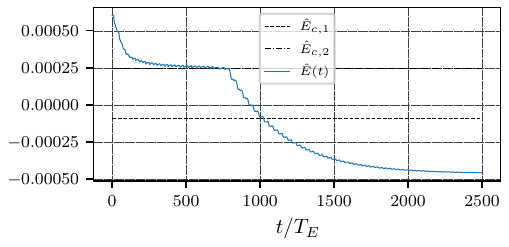}}
    \caption{Energy behavior of two cases from the first parameter study. On the left, $\kappa_{s}/\kappa^{*} = \kappa_{i}/\kappa^{*} = 0.8$, and the energy does not fall to the critical value needed for tidal locking. On the right, the large moment of inertia asymmetry $\kappa_{s}/\kappa^{*} = 6$ is sufficient to induce tidal locking, despite the sub-critical asymmetry in the interior, $\kappa_{i}/\kappa^{*} = 0.8$.}
    \label{fig:PSind1ab}
\end{figure}

It is instructive to examine two individual cases of interest from this parameter study. First, consider the non-synchronous rotation case with $\kappa_{s}/\kappa^{*} = \kappa_{i}/\kappa^{*} = 0.8$, whose energy plot is given on the left side of Figure \ref{fig:PSind1ab}. The energy settles to an equilibrium level above $\hat{E}_{\text{crit},2}$, and Europa ends up in a state of non-synchronous rotation. By contrast, for the case with $\kappa_{s}/\kappa^{*} = 6$ and $\kappa_{i}/\kappa^{*} = 0.8$, the energy falls to $\hat{E}_{\text{crit},2}$ and lingers until falling below this critical value, at which point tidal locking occurs, and energy is rapidly depleted afterwards. For this particular case, the ice shell and core tidally lock together when $\hat{E}(t) < \hat{E}_{\text{crit},2}$. It seems that regardless of the behavior of the ice shell, if the total energy falls below $E_{\text{crit,2}}$, tidal locking of the entire moon will eventually ensue, because equilibrium spin states with $E_{\text{crit},1} < E < E_{\text{crit},2}$ are generally not observed in any simulations for this first parameter study.

\begin{figure}[h!]
\centering
\includegraphics[]{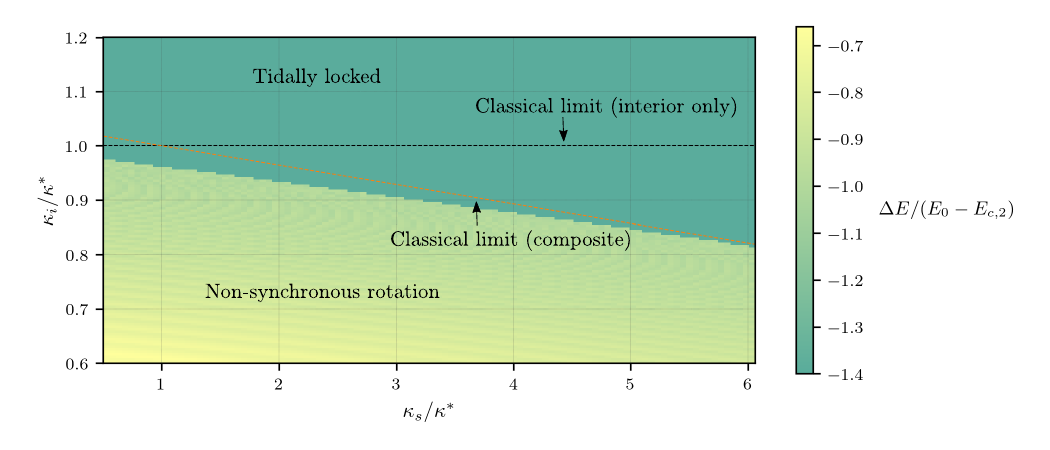}
\caption{Europa's equilibrium spin state vs. moment of inertia asymmetries $\kappa_{i}/\kappa^{*}$ and $\kappa_{s}/\kappa^{*}$, classical MacDonald torque, strong gravitational coupling case with $\hat{K}_{G}=1.6(B_{i}-A_{i})$ and $\gamma=28$. The figure shows $\Delta E/(E_{0}-E_{\text{crit},2})$, where $\Delta E = E_{f} - E_{0}$ is the energy decrease over 2500 orbits, and the dark cyan cases with $\Delta E/(E_{0}-E_{\text{crit},2}) < -1$ are all tidally locked. As opposed to the prior study with weak gravitational coupling, the boundary between tidal locking and non-synchronous rotation is simple and nearly linear in $\kappa_{s}$. This result is obtained with the classical MacDonald tidal torque. Over the same range of parameters, the modified tidal torque given by Williams \& Efroimsky results in universal tidal locking in this study.}
\label{fig:TLParam2}
\end{figure}

If the gravitational coupling between the permanent asymmetries of shell and interior is strong enough, the resulting behavior is similar to the classical result for a homogeneous satellite. Repeating the previous parameter study with the same methodology and the same $\gamma=28$ but using a strong gravitational coupling constant $\hat{K}_{G} = 1.6(B_{i}-A_{i})$, $\hat{K}_{G} > \hat{K}_{G}^{*}$, Figure \ref{fig:TLParam2} is produced. For this system, the energy loss always enforces simultaneous locking of the initially co-rotating ice shell and core -- see the lower half of Figure~\ref{fig:Energies}. The result is a less interesting landscape of possible solutions, with a fairly linear boundary separating tidal locking from NSR final states. Note however that once again the linear extrapolation, which assumes a differentiated but rigid Europa, does not correctly reproduce the slope of the boundary line. It provides a decent approximation of the boundary but once again is incorrect in the region with lower $\kappa_{s}/\kappa^{*}$. 

\begin{figure}[h!]
\centering
\includegraphics[]{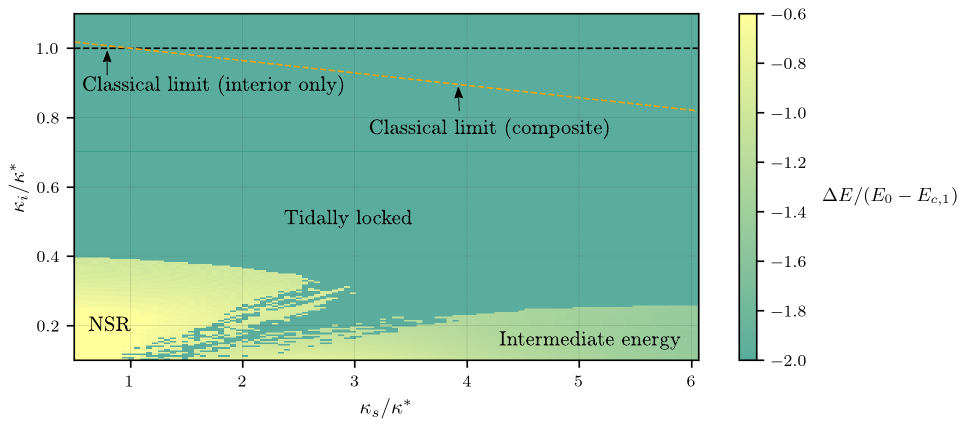}
\caption{Europa's equilibrium spin state vs. moment of inertia asymmetries $\kappa_{i}/\kappa^{*}$ and $\kappa_{s}/\kappa^{*}$, modified tidal torque, weak gravitational coupling case with $\hat{K}_{G}=0.6(B_{i}-A_{i})$ and $\gamma=28$. The value $\kappa^{*} = \frac{2}{3}\left(\frac{9.5\pi e^{2}}{2\sqrt{2}}\right)^{2}$ is the tidal locking $(B-A)/C$ limit from \cite{Goldreich_TAJ1966}, obtained using the classical MacDonald tidal torque. In the figure, the final energy determines the equilibrium state. This result is obtained with the modified MacDonald tidal torque of \cite{EfroimskyTidal}. In comparison to the result with the classical MacDonald tide, the region of NSR is drastically reduced to much lower values of $\kappa_{i}$.}
\label{fig:TLParamEfr}
\end{figure}

\begin{figure}[h!]
    \centering
    \subfloat[Normal ZVC, $\kappa_{s}/\kappa_{i} = 4/0.38 = 10.5$]{\includegraphics[scale=0.85]{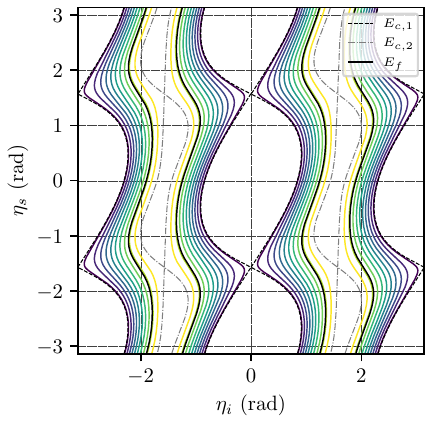}} \ \ \ \ 
    \subfloat[Modified ZVC, $\kappa_{s}/\kappa_{i} = 4/0.38 = 18.2$]{\includegraphics[scale=0.85]{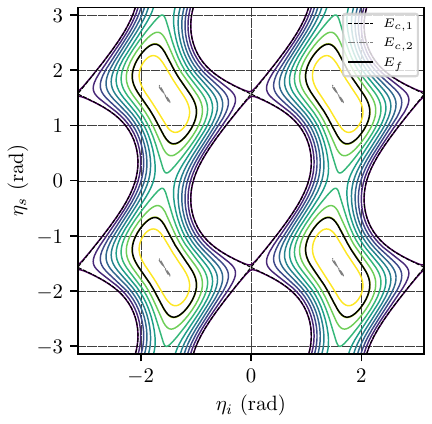}}
    \caption{Zero-velocity curves characteristically modified by a sufficiently large $\kappa_{s}/\kappa_{i}$. In both panels, the zero-velocity curves for the two critical energies are given, along with curves for increasing energies (with warming colors) from $E_{\text{crit},1}$ to $E_{\text{crit},2}$. In panel (a), the zero-velocity curves are shown to be analogous to those in Figure \ref{fig:ZVC_final}, though deformed, whereas in panel (b), $E_{\text{crit},2}$ no longer serves as a meaningful boundary. For both cases we also give a sample final energy $E_{f}$ which is just slightly less than $E_{\text{crit},2}$, corresponding to the bounds of some intermediate-energy state.}
    \label{fig:ZVC_changed}
\end{figure}

Repeating the parameter study that produced Figure~\ref{fig:TLParam}, but using the alternate tidal model introduced by \cite{EfroimskyTidal}, we find that much lower levels of permanent moment of inertia asymmetry are required for NSR to be permitted with this alternate tidal model. Expanding the range of $\kappa_{i}/\kappa^{*}$ from $[0.6, 1.2]$ to $[0.1, 1.2]$, and retaining the same range of $\kappa_{s}$, Figure \ref{fig:TLParamEfr} is produced. The result is a drastic reduction in the range of NSR to very low values of $\kappa_{i}/\kappa^{*}$. Additionally, the complex region of mixed NSR and tidal locking solutions is shifted to lower values of $\kappa_{s}/\kappa^{*}$ in comparison to the result obtained with the classical MacDonald tidal torque, in Figure~\ref{fig:TLParam}. This region extends so low that even $\kappa_{s} = \kappa^{*}$ is potentially low enough to induce tidal locking, if the interior MoI asymmetry is very small. 

In Figure \ref{fig:TLParamEfr} we note final spin states of intermediate energy $E_{\text{crit},1} < E_{f} < E_{\text{crit},2}$ in the bottom-right of the plot, not seen previously. This is for cases of very small $\kappa_{i}$ and very large $\kappa_{s}$. An analysis of these new cases reveals end states where the interior is undergoing NSR but the ice shell is tidally locked. This surprising result, not otherwise seen in our simulations, can be explained by revisiting the zero-velocity curves as shown in Figures~\ref{fig:ZVC_final} and \ref{fig:Energies} for the case of $\gamma = C_{i}/C_{s} = 28$ and $\kappa_{i} = \kappa_{s}$. It turns out that if the value of $\kappa_{s}/\kappa_{i}$ gets sufficiently large, those conclusions need to be modified. Consider the two sets of zero-velocity curves given in Figure~\ref{fig:ZVC_changed} for values of $\kappa_{s}/\kappa_{i} = 4/0.38 = 10.5$ and $\kappa_{s}/\kappa_{i} = 4/0.38 = 18.2$. In the transition between these two, the zero-velocity curves change in a way that drastically changes the qualitative limitations on spin evolution.
In particular, for the latter, the interior is still allowed to revolve at intermediate energy, even if the shell tidally locks. 

\FloatBarrier 
\subsection{Other Torques on Europa's Ice Shell}
Here, to aid in understanding the limitations of our present analysis, we provide a summary of the most commonly discussed significant torques that could also be at work within Europa during general independent motion of the shell and interior. Some of these torques are conservative, and some are dissipative. In the case that a torque is conservative, a new energy function similar to the one introduced in this paper can be developed to account for its effects, via Eqs.~\eqref{VfromL1} and \eqref{GeneralHam}. In the case that a torque is dissipative, there is a corresponding decrease in total system energy due to the action of the dissipating torque. The dissipation power is the product of the dissipative torques and the instantaneous angular velocity of the associated body, as shown in Eq.~\eqref{EnergyDot}.

The first torque to consider is the oceanic pressure torque, which consists of both the hydrostatic pressure torque and the Poincaré torque. 
Consider the general torque induced by fluid pressure on a surface of body $j$:
\begin{equation}
\label{PressureTorque1}
\bm{L}_{\text{p},j} = \int_{S_{j}}P \left(\bm{r} \times \hat{\bm{n}}\right) \ \text{d}S_{j}
\end{equation}
where $P$ is the pressure exerted by the fluid at point $\bm{r}$ on the surface, $\hat{\bm{n}}$ is the local normal vector of the surface of body $j$, and the expression is integrated over the entire surface $S_{j}$. The Van Hoolst model, which was the primary focus of this work, neglects the possibility for Europa's ocean to directly influence the ice shell rotational dynamics, and it is important to examine this assumption. 
The Poincaré torque is exerted when the polar axis of a rotating fluid departs from the axis of figure of the enclosing shell. In the case of Europa, this term is thus important when the axis of figure of the enclosing shell departs significantly from Europa's rotation axis. This torque is conservative, but should be negligible for our NSR analysis which is only about the $z$ axis. It is sometimes necessary to consider this term in a polar wander analysis -- where the spin axis and axis of figure depart significantly \citep{Ojakangas1989EuropaPW}. Neglecting the unimportant Poincaré torque, to provide a simplified look at the influence of a dynamically relevant ocean on Europa's rotation, consider the approach from \cite{BalandVanHoolst}, which modifies Van Hoolst's original model by adding oceanic pressure torques calculated from a simple hydrostatic balance. This analysis assumes that the fluid pressure obeys the (inertial) Navier-Stokes hydrodynamical equation,
\begin{equation}
\label{NStokes1}
\nabla\left(\frac{P}{\rho} - \Phi_{0} - W\right) = \bm{0}
\end{equation}
for pressure at a point $P$, density $\rho$, self-gravitational potential $\Phi_{0}$, and disturbing (Jovian) gravitational potential $W$. Note this is written as seen in a non-rotating frame, and there should be a small additional contribution of a transport term if the analysis is undertaken in a frame co-rotating with Europa's orbit. Using Gauss's theorem, the pressure torque can be computed  via the following volume integral based on the gradient of the internal gravitational potential and the Jovian gravitational disturbance:
\begin{equation}
\label{PressureTorque2}
\bm{L}_{\text{p},j} =  \int_{V_{j}}\bm{r}\times \nabla P(\bm{r})\text{d}V_{j} = \int_{V_{j}}\bm{r}\times \rho\left(\nabla\Phi_{0}(\bm{r}) + \nabla W(\bm{r})\right)\text{d}V_{j}
\end{equation}
The resulting pressure torque on Europa's ice shell will be strongly dictated by the global shape of Europa, which is dominated by the hydrostatic tidal response of the solid interior and the global ocean and floating ice shell \citep{Nimmo2007TheGS}. The result in \cite{BalandVanHoolst} gives coupling pressure torques for a global subsurface ocean for a differentiated icy Galilean satellite. However, the results presented for that librational work are computed specifically for angles between the hydrostatic ellipsoidal figures, whereas for this work we'd need to compute the influence of pressure torques induced by topographic differences secondary to the static tidal figures. Such a calculation is beyond the scope of this work. It might be necessary to revisit for satellites far from their primary like Callisto or Titan, where the hydrostatic bulge is expected to be less pronounced, making the topographic contribution comparatively greater. Beyond this, non-hydrostatic effects could be quite important to the net overall oceanic torque contribution, having been demonstrated as a possibly dominant oceanic effect in Titan's length-of-day variations \citep{VanHoolst_Titan2009}, for example.

On the topic of non-hydrostatic effects, the friction torque is another oceanic torque to consider, due to the differential angular velocity of the ice shell and interior. It is a dissipative torque, which has previously been assumed to be turbulent \citep{Ojakangas1988}:
\begin{equation}
\label{FrictionTorque2}
\bm{L}_{\text{Turbulent}} \approx C_{T}\|\bm{\omega}_{i} - \bm{\omega}_{s}\|(\bm{\omega}_{i} - \bm{\omega}_{s})
\end{equation}
where $C_{T} \propto r^{5}$ for radius $r$. The condition for turbulence is $\text{Re} = \frac{ul}{\nu} \gg 300$, and a reorientation timescale of $10^{5} - 10^{6}$ yrs would yield $\text{Re} \sim 10^{5} - 10^{6}$, greatly exceeding the level required for turbulence. \cite{Ojakangas1988} argues that even a more powerful laminar friction torque would still likely be sub-dominant. However, because the friction torque scales with the ice shell reorientation rate, it would act to impede ice shell reorientations above some maximal rate.
The net torque effects of the ocean are complex and still debated. New findings suggest that depending on the strength of ocean convection, the ocean could exert torques that assist the spinning up or spinning down of the ice shell, primarily via friction drag at the ocean-ice shell boundary \citep{HamishHay2023}. This effect can only be studied via high-fidelity finite volume discretization models such as \cite{OceansFEM}, well beyond the scope or goals of this work. Given the uncertainty of the actual fluid torques on the ice shell, we defer in-depth consideration of the action of the ocean to future work.

The final torque under consideration is the shell dissipation torque. Because the ice shell is not perfectly rigid, there will be some loss of energy due to viscous flow, and potentially shell fracturing. This loss of energy due to work done on the shell implies that there would be a corresponding net torque on the ice shell. \cite{Ojakangas1989EuropaPW} gives a rough estimate of this torque for polar wander, assuming a continuous viscous shell and neglecting the effects of deep fractures and other geologically developed structural irregularities. Their expression is adapted for the NSR analysis with general shell and interior angular velocities $\bm{\omega}_{s}$ and $\bm{\omega}_{i}$: 
\begin{equation}
\label{ViscousTorque1}
\bm{L}_{v} \approx -C_{v}\frac{(\bm{\omega}_{s} - \bm{\omega}_{i})}{\|\bm{\omega}_{s} - \bm{\omega}_{i}\|}
\end{equation}
where $C_{v} \sim \frac{4\pi a^{2}d}{l}\left(\frac{T_{m}}{T_{m}-T_{s}}\right)\mu \gamma^{2}$ for average radius $a$, ice depth $d$, melting temperature $T_{m}$, characteristic surface temperature $T_{s}$, elastic modulus $\mu$, and a constant $l$. This is derived by a simplified model dividing the ice shell into an elastically responding outer layer and viscously responding inner layer. Ojakangas shows that this estimate for the torque is massive, and discusses at length the difficulty of accurately estimating the scale of viscous dissipation in a realistic model of the shell. In particular, this dissipative torque could be greatly reduced by deep fractures in the ice, with the strain accommodated by the cracks between fractured elements of the shell instead of in the shell elements themselves. However it is not known to what extent and to what depth Europa's ice shell has historically been fractured.

An appropriate extension of this work should include the dissipative torque associated with the shell's viscoelastic nature, and could also include the (likely turbulent) friction torques, but the dissipation due to shell non-rigidity is likely the dominant of the two. Because this effect removes a lot of energy from the system, we would expect the NSR boundaries in Figures ~\ref{fig:TLParam} and \ref{fig:TLParam2} to shift. Intuitively we'd expect that the regions of tidal locking will generally encroach on the regions of non-synchronous rotation due to the presence of an additional avenue for energy dissipation below the critical threshold for tidal locking. However, these additional coupling torques might also impede the development of the independent oscillations of the shell and core observed in this work. This would result in a greater portion of the total system energy remaining in the rotation of the solid interior, and potentially act as a buffer against the chaotic shell rotation solutions which feature tidal locking of the interior.

\subsection{Non-Rigidity in Europa's Early Ice Shell}
While we leave numerical exploration of the response of a non-rigid ice shell to future work, the implications of the effects of non-rigidity are still a valuable discussion, especially in the context of the kinds of motions predicted by the analyses in this work. \cite{Ojakangas1989EuropaPW} discusses the damping effects of viscous dissipation for a non-rigid ice shell that undergoes rapid angular displacement, showing that viscous effects can absorb a large amount of energy, significantly impeding rapid shell motion. The decoupled shell revolution behavior in our work may be stalled by dissipation, or alternatively a massive amount of energy is quickly imparted into the shell -- potentially fracturing the ice. The type of reaction depends on the thickness of the ice shell. Since we are exploring tidal locking which could have occurred at any point in Europa's history, the effect on an ice shell much thinner than it is today (representative of an earlier warmer Europa) is also of interest. 

The Maxwell time $\tau_{m}$ generally decreases with depth in the ice shell. Consider also a ``reorientation time" $\tau_{r}$ representative of the timescale for the ice shell to revolve to a longitude 90 degrees offset from that of the interior. For a primordial thin ice shell, the response is dominated by a large $\tau_{m}$. If the reorientation time is significantly under this timescale, $\tau_{r} \ll \tau_{m}$, it could induce broad global fracturing in the shell due to the inability for elastic strain to compensate the entire deformation. Thus, our solutions that forced the shell into a chaotic spin state (characterized by independent revolutions from the core) could in reality impart a significant amount of energy into the broad fracturing of the shell. Such a geologically violent event could also drastically change the scale and location of the permanent mass asymmetries in the shell, altering $\kappa_{s}$ and additionally defining a new $\eta_{s}$. Depending on the amount of energy released in a cracking incident, such a mechanism of inducing global cracking could present a means of dissipating rotational energy throughout Europa's early history, as the shell cracks, re-freezes, and rotates out of alignment from the core repeatedly. However, note that the chaotic shell rotation solution was only observed for part of the parameter space, in particular for some solutions where $\kappa_{i} < \kappa^{*}$ and $\kappa_{s} \gg \kappa^{*}$. These are cases when the interior has a small mass asymmetry but the shell has a large asymmetry. 
\begin{figure}[h!]
\centering
\subfloat[Thin, fractured ice shell]{\includegraphics[width=2.5in]{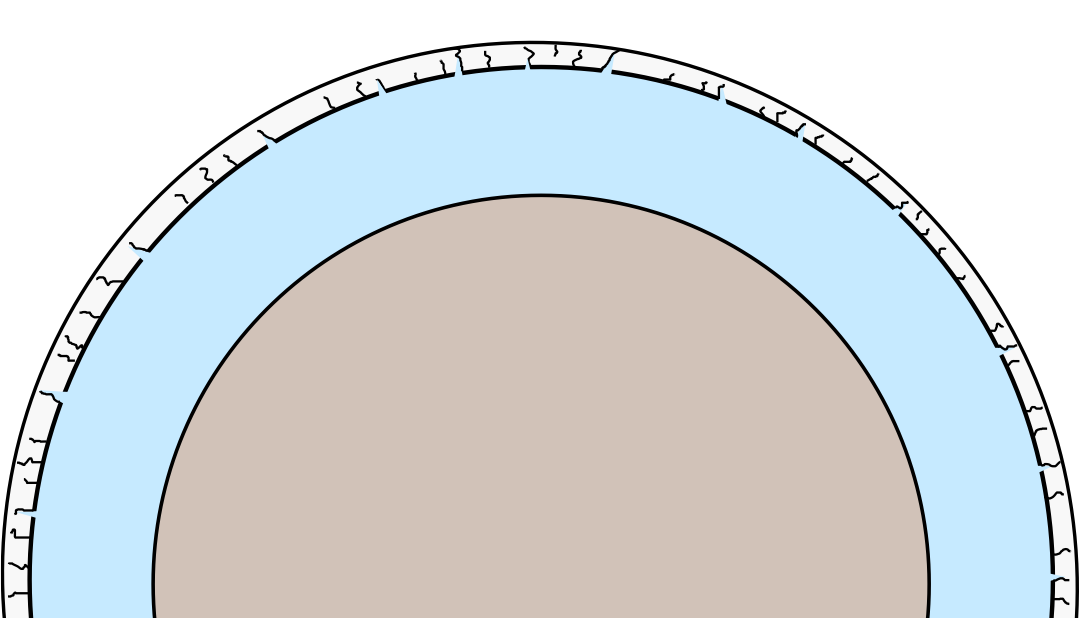}} \hspace*{+2.0em}
\subfloat[Thick, dissipative ice shell]{\includegraphics[width=2.5in]{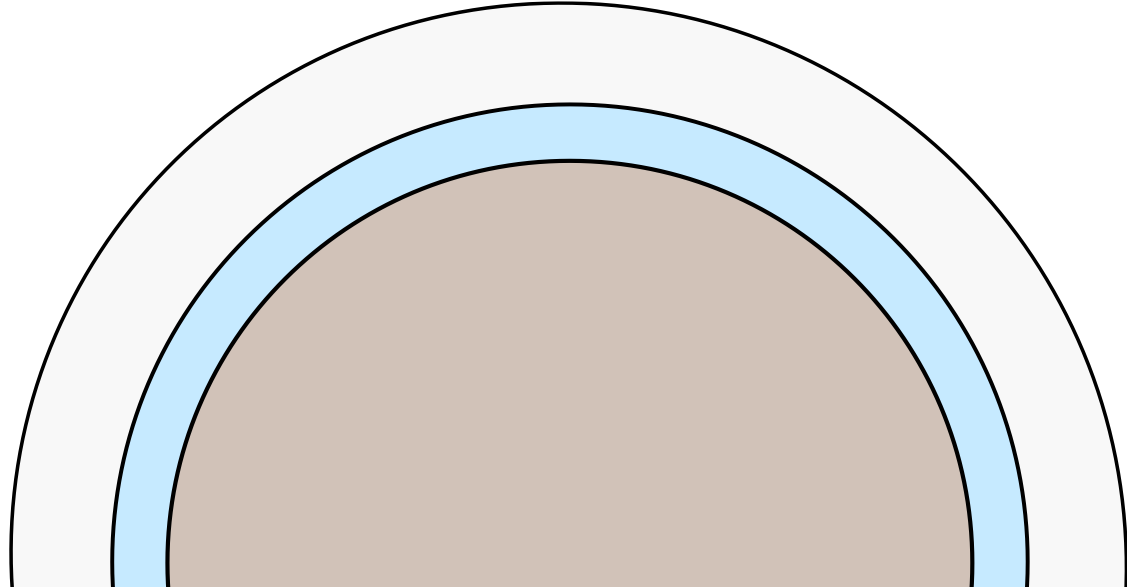}}
\caption{Two qualitatively different responses of Europa's ice shell to a large independent longitudinal displacement. On the left is Europa with a thin, fractured shell. The reorientation timescale is significantly less than the Maxwell time, $\tau_{r} \ll \tau_{m}$, and the rheological response to reorientation-induced deformation is broad brittle fracturing. On the right is Europa with a thick, dissipative shell. The reorientation timescale is significantly greater than the Maxwell time, $\tau_{r} \gg \tau_{m}$, and the response to large angular displacement is viscous dissipation.}
\end{figure}

In an alternate scenario, consider a thick ice shell whose response is dominated by the small $\tau_{m}$ of the warm ice deep in the shell, extending down to the shell-ocean interface. For this thick shell, if the reorientation timescale satisfies $\tau_{r} \gg\tau_{m}$, there could be a rapid arresting of independent motion due to dissipation deep within the ice shell. However, deep cracks in the ice could act to reduce this dissipative effect. The degree to which dissipation would be reduced is not certain, because an accurate modeling of the global viscoelastic response of the shell is extremely complicated, and the shell fracturing vs. depth has not been characterized. 

\subsection{Mass Asymmetries of Europa's Shell and Interior}
Mass asymmetries in Europa's icy shell and interior may exist due to variations in topography and density, with potentially significant contributions to the moment difference $\kappa = (B - A)/C$. In fact, only the ``frozen-in" asymmetries contribute to $\kappa$, because the hydrostatic figure raised by tides is directly responsible for nonzero mean torque driving non-synchronous rotation, whereas $\kappa$ controls the restoring torque when the asymmetry is misaligned with the tidal bulge \citep{Goldreich_Peale_SpinOrb}. In the context of the dynamical coupling of the shell and interior, the respective values of $\kappa_{s}$ and $\kappa_{i}$ are thus proportional to the quadrupole moments, $l = m = 2$ of the shell and interior mass distribution, where $C_{lm}$ are the resulting spherical harmonic coefficients for the gravity field of the rigid body.

First, we consider small-scale topography. For reasonable values of Europa's ice shell thickness $d$, a circular load with diameter $L$ will remain rigidly supported for timescales $t_\mathrm{R} \sim \eta/\rho g L$, where $\eta$ is the shell's viscosity (Pa s), $\rho$ is the density, and $g$ is the surface gravity. The temperature-dependence of ice viscosity follows an Arrhenius-like relation, 
    $\eta(T) = \eta_0 \exp{\left\{\frac{A_l}{R_\mathrm{gas}}\left(\frac{1}{T}-\frac{1}{T_\mathrm{m}}\right)\right\}}$
where $\eta_0 \approx 2\times 10^{14}~\mathrm{Pa~s}$ is the viscosity at the melting point $T_\mathrm{m} \approx 270~\mathrm{K}$, $A_l \approx 6\times 10^4~\mathrm{J~mol^{-1}}$ is the activation energy for material diffusion, and $R_\mathrm{gas} = 8.3145~\mathrm{J~mol^{-1}~K^{-1}}$ is the universal gas constant \citep{ashkenazy2018dynamics}. Given that the maximum stress beneath a surface load occurs at a depth $z_\mathrm{max} \sim L/3$, with these ice properties and a geothermal gradient $dT/dz = 10~\mathrm{K~km^{-1}}$, we find $t_\mathrm{R} > 10^7~\mathrm{yr}$ (Europa's approximate present-day surface age) for $L < 20~\mathrm{km}$. In other words, topographic loads with horizontal dimension smaller than roughly 15 km will be supported by the rigidity of the upper part of the ice shell; broader loads will be supported by isostatic compensation at the base of the shell. Compensation at depth $d$ reduces the net contribution of a load to $(B-A)/C$ by a factor $d/R \sim 0.01$ \citep{Greenberg1984}. A similar analysis applies to the rocky interior, but we lack constraints on its rheological properties. Taking a viscosity $\eta_i \sim 10^{21}~\mathrm{Pa~s}$ (comparable to Earth's mantle), we find similar-sized loads as above, $L \sim 10-20~\mathrm{km}$ are likely to be rigidly supported by the rocky interior. 

The mass of the anomaly is $\Delta m \sim L^2\Delta z \rho$, where $\Delta z$ is the topographic anomaly relative to the ellipsoidal surface. Although topography data for Europa are limited, limb profiles suggest variations of $\Delta z \sim 1~\mathrm{km}$ over horizontal baselines of $\sim 10^\circ = 270~\mathrm{km}$ \citep{Nimmo2007TheGS}, while relief across chaos and ridged plains regions was measured to be $\sim 100~\mathrm{m}$ on 10-km baselines \citep{SchenkPappalardo2004}. Therefore, for $L = 20~\mathrm{km}$, we assume $\Delta z \sim 200~\mathrm{m}$, yielding $\Delta m \sim 10^{14}~\mathrm{kg}$. If the mass anomaly is positioned at the sub-Jovian point at the equator, the effective change to the shell's mass asymmetry is $\Delta (B-A)/C \sim \Delta m / m_s \sim (10^{14}~\mathrm{kg})/(10^{22}~\mathrm{kg}) \sim 10^{-8}$. Given the required (classical) value $\kappa^* \sim 10^{-6}$ \citep{Greenberg1984}, this result indicates Europa's rigidly supported topography is $\sim 100\times$ smaller than needed to achieve the critical value. Again, a similar analysis applies to the rocky interior, but the effects of topography are further reduced, due to the much larger value of the polar moment of inertia $C_i$.

In addition to local mass anomalies, $(B - A)/C$ may be affected by global-scale asymmetries in the shell thickness, or for the interior, a frozen-in ellipsoidal bulge. Thickness variations in Europa's ice shell would affect $\kappa_s$. \cite{Ojakangas1989EuropaThermal} showed that
\begin{equation}
    \Delta \left(\frac{B-A}{C}\right) \sim \left(\frac{\rho_\mathrm{m}-\rho_\mathrm{c}}{\overline{\rho}}\right)\left( \frac{\rho_\mathrm{c}}{\rho_\mathrm{m}}\right)\left(\frac{d}{R}\right)\left( \frac{4t_{22}}{R}\right),
\end{equation}
where $\rho_\mathrm{m}$ and $\rho_\mathrm{c}$ are the density of the liquid water `mantle' and icy shell, respectively, $R$ is Europa's mean radius, $d$ is the mean shell thickness, and $t_{lm}$ represents the topography contained in the $(l,m)$ spherical harmonic. Using limb profiles, \cite{Nimmo2007TheGS} found a limit to maximum global thickness variations that requires $t_{22} < 1~\mathrm{km}$, or $4t_{22}/R \sim 2\times 10^{-3}$. Therefore, with $\rho_\mathrm{m} = 1000~\mathrm{kg~m^{-3}}$, $\rho_\mathrm{c} = 920~\mathrm{kg~m^{-3}}$, $\overline{\rho} = 990~\mathrm{kg}$, $d = 20~\mathrm{km}$, $R = 1560~\mathrm{km}$, we find $\Delta (B-A)/C \lesssim 10^{-6}$. Thus, shell thickness variations and associated topography may be barely too small to cause $\kappa_s$ to exceed the classical $\kappa^*$ computed using Goldreich and Peale. However, these shell thickness variations would be large enough to exceed the comparatively smaller $\kappa^*$ obtained from the tidal model of \cite{EfroimskyTidal}. 

The shape of the silicate interior is unknown, either in the past or present. We can make comparisons to similar-sized silicate bodies, such as Io and Earth's Moon. Topography data from the Lunar Orbiter Laser Altimeter (LOLA) (Mazarico et al., 2013) and limb profiles of Io \citep{White2014_IoLimb} indicated almost identical values of $4t_{22}/R \sim 2\times 10^{-4}$. Assuming a silicate crustal density $\rho_\mathrm{c} = 3000~\mathrm{kg~m^{-3}}$ and mantle density $\rho_\mathrm{m} = 3500$ (similar to Io's mean density), with a 1 km-thick crustal layer, we find $\Delta (B-A)/C \sim 10^{-8}$. Thus, if the frozen-in shape of Europa's silicate core were similar to Io's or the Moon's, the resulting $\kappa_i$ would be much smaller than $\kappa^*$. However, it should be noted that if Europa's interior were volcanically active throughout its early history, much larger asymmetries are possible. For example, Mars has $4t_{22}/R \sim 10^{-2}$, due to the Tharsis rise; an asymmetry of similar relative thickness on Europa's seafloor would result in $\kappa_i > \kappa^*$ for both Goldreich's classical $\kappa^{*}$ and the smaller limit using the modified tidal model.

\section{Discussion and Conclusions}

In this work, we explore the tidal locking process for Europa, differentiated into a solid interior, ocean layer, and ice shell. Classically, as a planet's rotation slows due to energy loss, the ``dynamic" tidal torque, induced by non-rigidity of the satellite, acts to produce an equilibrium spin state slightly faster than the synchronous angular velocity. This effect is countered by the ``static" torques acting on permanent asymmetries in the body. Our work explores how differentiation of a moon into a gravitationally interacting solid interior and ice shell affects classical conclusions about the evolution of its rotational state. First, we note that the equations of motion previously studied \citep{VanHoolst_Europa} for rigid shell and core interacting by mutual gravitation admit an energy integral. This quantity changes slowly under the action of unmodeled perturbative torques such as the dynamic tidal torque. We find that the spin states of the shell and core are constrained by the instantaneous value of the system's energy state, with critical values of the energy demarcating qualitatively distinct behaviors. Furthermore, the dynamical evolution of the spin state with energy loss is governed by the strength of gravitational coupling between the ice shell and solid interior. Below a critical value of gravitational coupling, the ice shell can revolve independently of the solid interior, and tidal locking of the solid interior precedes an era of erratic independent shell rotation. This behavior is much more common for certain combinations of ice shell and interior moment of inertia asymmetries. Above the critical value of gravitational coupling, the ice shell and interior revolve and tidally lock in lockstep. 

Numerical simulation confirms the validity of using the system energy as a lens for studying and describing overall behavior. We show that the 2D (four state) dynamics of the model are chaotic outside of the region of librations and synchronized revolutions of the shell and interior. We provide several case studies adding the effects of orbital eccentricity and dynamic tidal torque, demonstrating examples of tidal locking for strongly and weakly gravitationally coupled cases, as well as examples showing non-synchronous rotation as a final spin state. We note that the assumptions made for modeling the dynamic tidal torque can greatly affect the resulting final spin state. In particular, use of the alternate model introduced by \cite{EfroimskyTidal} results in a far higher likelihood for tidal locking than use of the classical MacDonald model. With our higher fidelity models, we execute large parameter studies over a range of normalized asymmetries $\kappa_{s}/\kappa^{*}$ and $\kappa_{i}/\kappa^{*}$ for both the shell and solid interior, where the normalization parameter $\kappa^{*}$ is the necessary moment of inertia asymmetry $(B - A)/C$ classically required for a rigid satellite to tidally lock \citep{Goldreich_TAJ1966}. Two studies are performed -- one for a gravitational coupling parameter less than its critical value $K_{G} < K_{G}^{*}$ and one greater, $K_{G} > K_{G}^{*}$. For both studies, Europa is subject to dynamic tidal torques high enough to bring the moon to its final spin state within a reasonable span of simulation time, similar to the procedure used in \cite{Goldreich_TAJ1966} for the 1D rigid satellite case. Using the energy parameter introduced in this paper, we efficiently differentiate tidally locked from non-synchronously rotating final spin states across the entire parameter space. Then, we compare the resulting distribution of final spin states to the results predicted by \cite{Goldreich_TAJ1966}  for a rigid satellite. We note that particularly for the case of weak gravitational coupling, the differentiation of shell and interior plays an important role in the dynamical evolution of Europa's spin state ways that are not captured by the rigid moon model. Finally, we discuss neglected torques that are expected to be sub-dominant, and provide commentary on the potential effects of non-rigidity of the ice shell. In the future, viscoelastic extensions of the model explored in this paper could provide additional realism and new insights. 

Our results highlight that the final spin state Europa experiences can greatly depend on which model is used for the dynamic tidal torque. 
Keeping in mind that the MacDonald torque-based classical expression given by Eq.~\eqref{Ean2} is mathematically inconsistent, we apply the model of \cite{EfroimskyTidal}, and we obtain a significant contraction of the boundary demarcating NSR from tidal locking. As discussed in \cite{MakarovEfroimsky2013}, use of a tidal model more reflective of realistic rheologies can result in the destruction of NSR as a stable equilibrium spin state altogether. Additionally, in this work we highlight other dissipative effects characteristic of a decoupled floating ice shell (namely ice shell global fracturing or viscous arrest, depending on shell thickness). From the perspective of the energy integral and critical energies and their relation to tidal locking, we argue that such additional avenues for dissipation could reduce equilibrium energy and thus contract the range of NSR solutions further. It is clear that the NSR result as explored by Goldreich \& Peale is highly sensitive to the choice of tidal model, and is obtained from first-order analysis that neglects potentially important dissipative paths of realistic rheologies. In reality, the extremely slow NSR that has previously been argued to potentially exist at Europa would have to persist in spite of these dissipative pathways. If the ice shell does indeed undergo slow longitudinal migration, a non-hydrostatic driver such as that recently highlighted by \cite{HamishHay2023} might be required.

\section*{Acknowledgements}
Part of this work was supported by the Europa Clipper project, National Aeronautics and Space Administration. 

\bibliographystyle{apalike}
\bibliography{references.bib}   

\end{document}